\documentclass[prb,aps,showpacs,twocolumn,preprintnumbers,amsmath,amssymb,superscriptaddress,longbibliography,nofootinbib]{revtex4-2}
\pdfoutput=1
\usepackage[english]{babel}
\usepackage{amsmath,amssymb,bbm,mathrsfs,mathtools,multirow,diagbox,xcolor,graphicx,tabularx,comment,amsfonts,dsfont,lettrine,units,comment}

\newcolumntype{L}{>{\centering\arraybackslash}X}

\usepackage{graphicx}
\usepackage[colorlinks=True,linkcolor=blue,citecolor=blue,urlcolor=blue]{hyperref}
\usepackage{bookmark}
\usepackage{xcolor}
\usepackage{chngcntr}
\usepackage{braket}
\usepackage{nicefrac}
\usepackage{bm}
\usepackage{bbm}
\usepackage{braket}
\usepackage{lineno}
\usepackage{array}
\newcolumntype{C}{>{$}c<{$}}
\usepackage{newtxtext} 
\usepackage[normalem]{ulem}
\usepackage{titletoc}

\titlecontents{section}[1.5em]{}{\contentslabel{1.5em}}{\hspace*{-2.3em}}{}

\renewcommand{\dag}{^{\dagger}}

\renewcommand{\dag}{^{\dagger}}

\usepackage{gensymb}

\begin{document}

\title{{Doping-induced} nematic and stripe orders within the charge density wave state of TiSe$_2$}

\author{Daniel Mu\~noz-Segovia}
\email{dm3886@columbia.edu}
\affiliation{Donostia International Physics Center, P. Manuel de Lardizabal 4, 20018 Donostia-San Sebastian, Spain}
\affiliation{{Department of Physics, Columbia University, New York, NY 10027, USA}}
\author{J\"orn W. F. Venderbos}
\affiliation{Department of Physics, Drexel University, Philadelphia, PA 19104, USA}
\affiliation{Department of Materials Science and Engineering, Drexel University, Philadelphia, PA 19104, USA}
\author{Adolfo G. Grushin}
\affiliation{Institut Ne\'el, CNRS and Universit\'e Grenoble Alpes, Grenoble, France}
\author{Fernando de Juan}
\email{fernando.dejuan@dipc.org}
\affiliation{Donostia International Physics Center, P. Manuel de Lardizabal 4, 20018 Donostia-San Sebastian, Spain}
\affiliation{IKERBASQUE, Basque Foundation for Science, Maria Diaz de Haro 3, 48013 Bilbao, Spain}

\date{\today}
\begin{abstract}
In this work, we present a theory to {address} conflicting experimental claims regarding the charge density wave (CDW) state in TiSe$_2$, including whether there is a single or multiple CDW transitions and {whether threefold rotation symmetry ($C_3$) is broken}. Using a {continuum} $\boldsymbol{k}\cdot\boldsymbol{p}$ model coupled to the CDW order parameter, we show how commonplace conduction band doping {induces a nematic transition from a $C_3$-symmetric $3Q$ CDW to a $C_3$-breaking $3Q$ CDW, which is favored by the large ellipticity of the conduction bands of TiSe$_2$. We also find that a $1Q$ stripe CDW is generically stabilized for sufficiently high electron doping.} We then show how both stripe and nematic {CDW} states emerge self-consistently from a minimal interacting tight-binding model, for both positive and negative initial gaps. Our theory {provides a new scenario in which, as temperature is lowered, a second $C_3$-breaking transition may occur or not depending on the doping level, potentially explaining the experimental variability. These predictions can be further verified with a variety of probes including transport, photoemission and tunneling.}
\end{abstract}
\maketitle

\section{Introduction}
\begin{figure}
    \centering
    \includegraphics[width=0.45\textwidth]{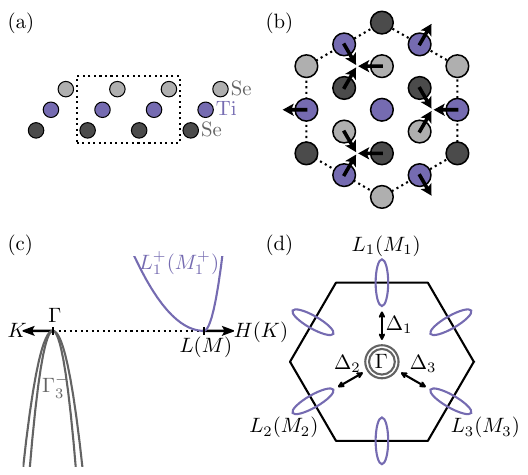}
    \caption{(a) Side view of the lattice structure of TiSe$_2$. 
    (b) Real space view of the charge density wave pattern. 
    (c) Low energy band structure and symmetry labels for the bands for the bulk (notation in parenthesis for the monolayer). 
    (d) Fermi surface sketch of the normal state in the semimetallic case. The order parameter $\vec{\Delta}$ coupling conduction and valence bands is also shown. }
    \label{Fig1}
\end{figure}

The transition metal dichalcogenide TiSe$_2$ develops a commensurate $2\times2\times2$ charge density wave (CDW) transition below $T_c \sim 200 \mathrm{K}$ (shown in Fig.~\ref{Fig1}(a,b)), which has been under scrutiny for decades~\cite{DMW76,Rossnagel11}. Featuring nearly energetically aligned electron and hole pockets at the Brillouin Zone (BZ) $L$ and $\Gamma$ points (see Fig.~\ref{Fig1}(c,d)), TiSe$_2$ is naturally unstable to modulations with momentum $Q=\Gamma L_i$ which cause a repulsion of the electron and hole bands. These can include both lattice modulations driven by electron-phonon coupling~\cite{WRC11,CM11}, charge modulations driven by excitonic correlations~\cite{monney_spontaneous_2009,Kogar17} or, most likely, a combination of the two~\cite{vanWezel2010,WezelSaxena,porer_non-thermal_2014,monney_revealing_2016}. 

\begin{figure*}
    \centering
    \includegraphics[width=0.85\textwidth]{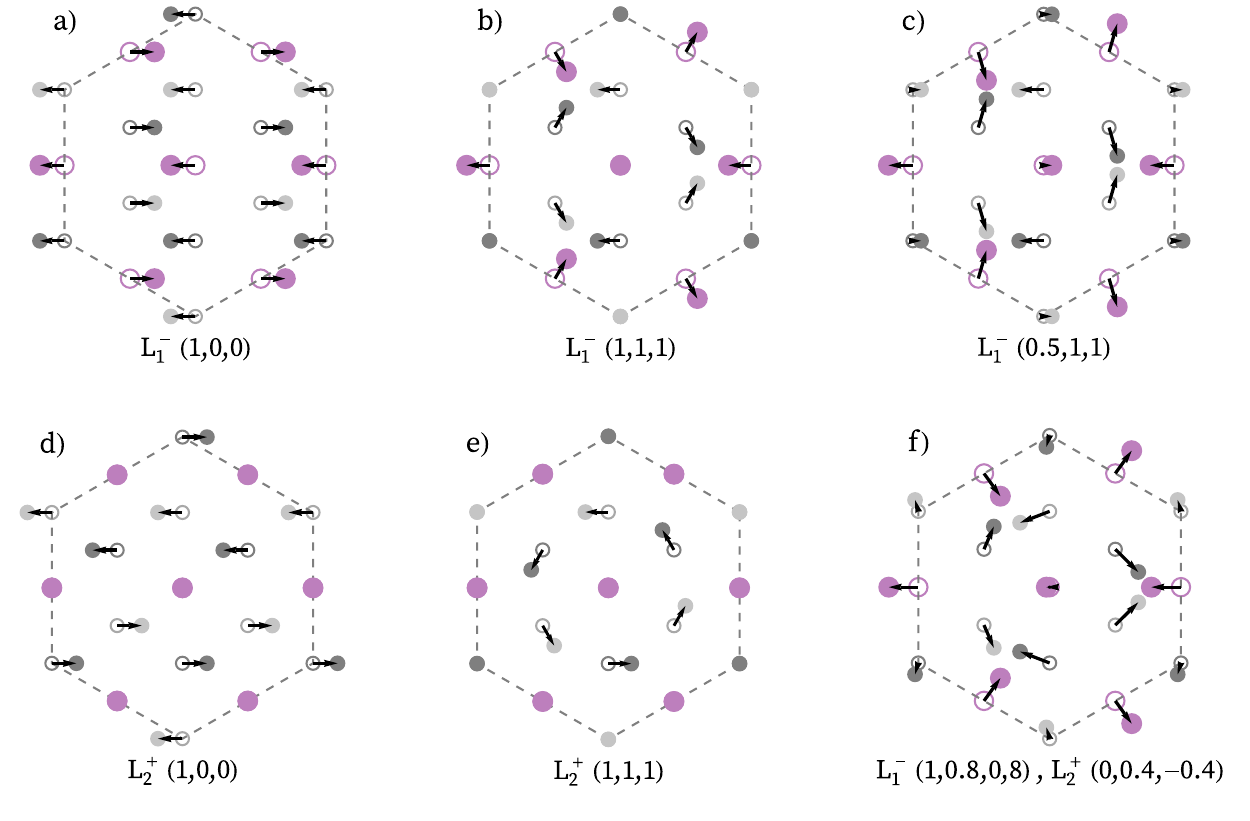}
    \caption{Atomic displacements for the different phonon configurations discussed in the text. Panels (a-c) represent the different states predicted in our work: $L_1^-$ phonon in $1Q$ stripe $(\Delta,0,0)$, $C_3$-symmetric $(\Delta,\Delta,\Delta)$, and nematic $3Q$ $(\Delta_1,\Delta_2,\Delta_2)$ configurations. (d,e) $L_2^-$ displacements in the $(\Phi,0,0)$ and $(\Phi,\Phi,\Phi)$ configurations. (f) Chiral CDW proposal from Ref. \cite{Peng22}, which involves the mixing of $L_1^-$ and $L_2^+$ configurations. Note that both (c) and (f) have a twofold symmetry axis $C_{2x}$. In an STM experiment which probes predominantly the upper Se lattice (dark gray), states (c) and (e) cannot be distinguished by symmetry.}
    \label{FigPhonons}
\end{figure*}

{While much effort has been devoted to identifying the microscopic origin of the CDW, a more pressing controversy regarding the symmetry of the ordered state remains unresolved}. The CDW order parameter $\vec \Delta$, which has three components representing the three $Q$ modulation vectors, was established by neutron diffraction \cite{DMW76} and X-ray \cite{Holt01} experiments to have $L_1^-$ symmetry, {which preserves an inversion center between Ti layers}. These experiments further showed that $\vec \Delta$ condenses in the $C_3$-symmetric $3Q$ configuration $\vec \Delta = (\Delta,\Delta,\Delta)$, {schematically shown in Figs.~\ref{Fig1}(b) and~\ref{FigPhonons}(b)}. Scanning tunneling microscopy (STM), on the other hand, observed {different intensities for the three CDW Bragg peaks at the surface BZ $\overline{M}_i$ points~\cite{ILS10}. Once other artifacts such as tip anisotropy have been discarded~\cite{Silva13}, this implies that $C_3$ symmetry has been spontaneously broken, as observed also in Kagome superconductors~\cite{Jiang21,LiNematic2022,Nie22} or twisted heterostructures~\cite{Jiang19,RubioVerdu22}. Inspired by the STM experiment of Ref.~\cite{ILS10} in TiSe$_2$}, a chiral CDW breaking all mirrors, inversion and $C_3$ was proposed~\cite{vanWezel11,van2012chiral,Zenker13,GW15,Peng22}. {While further STM works ~\cite{Ishioka11,Iavarone12,Kim24seamless} provided support to this picture, others clearly displayed a $C_3$-symmetric state~\cite{Hildebrand18}. Many other experiments~\cite{Castellan13,Lin19,Ueda21,Xiao24,Ueda25,ZhangSHG22,Tyulnev24,Xu20,jog_optically_2023}, reviewed in detail in Sec.~\ref{subsec:intro_sym_break}, have since argued in favor or against the chiral CDW hypothesis, thus reinvigorating the controversy.} 

{In this work, we propose a minimal theory to explain the $C_3$-breaking order and its experimental variability from sample to sample which, in contrast to the chiral CDW proposal, invokes only the condensation of the standard $L_1^-$ order parameter, which preserves inversion symmetry.} {Our symmetry analysis reveals that, if $ \vec \Delta$ condenses in the $C_3$-breaking form $(\Delta_1,\Delta_2,\Delta_2)$, which we call the \textit{nematic 3Q CDW} (see Fig.~\ref{FigPhonons}(c)), this is enough to explain the STM experiments~\cite{ILS10,Ishioka11,Iavarone12,Kim24seamless}.}

{Furthermore, we propose that the experimental variability may originate from the doping dependence of CDW state. Constructing a microscopic model, we show that electron doping of the CDW indeed drives further electronic instabilities to $C_3$-breaking CDW states, first in the form of the proposed nematic $3Q$ CDW state and then as a $1Q$ stripe CDW at larger doping (see Fig.~\ref{FigPhonons}(a)).} {Experimentally, it is known that the conduction band in TiSe$_2$ is often occupied by a small electron density, and achieving fully compensated, stoichiometric samples requires a careful growth method only described recently ~\cite{Campbell19,WBK19,Knowles20}. Control of the doping level is therefore key to addressing the variability of symmetries observed. Our theory makes testable predictions for the doping dependence of symmetry breaking pattern, and can thus help discern whether invoking new order parameters is required to explain experiments. Our theory pertains to the low-doping case only, $x \lesssim 0.06$ electrons per formula unit (e/f.u.), as higher levels of doping are known to first render the CDW incommensurate~\cite{KogarICDW,Yan17} and then give rise to superconductivity~\cite{Morosan06,LOL15} as predicted theoretically~\cite{Wei17,guster_first_2018,chen_reproduction_2018,Pereira19,Novko22}.}

{This paper is organized as follows. Section~\ref{sec:phenomenology} reviews the known phenomenology of TiSe$_2$ with a focus on symmetry analysis. We compare the nematic $3Q$ state proposed here and the chiral CDW proposal~\cite{vanWezel11,van2012chiral,Zenker13,GW15,Peng22}, and discuss them in the context of STM and other symmetry-sensitive experiments~\cite{ILS10,Ishioka11,Iavarone12,Hildebrand18,Kim24seamless,Castellan13,Lin19,Ueda21,Xiao24,Ueda25,ZhangSHG22,Tyulnev24,Xu20,jog_optically_2023}. We also emphasize the uncontrolled native doping present in the majority of the TiSe$_2$ samples synthesized to date~\cite{Watson19,Watson20,Campbell19}. In Section~\ref{sec:kp} we use the symmetries of TiSe$_2$ to construct a low-energy 2D $\boldsymbol{k}\cdot\boldsymbol{p}$ model coupled to the CDW $L_1^-$ order parameter. We show that the energetics of the electronic bands in the CDW state generically drives transitions from the $C_3$-symmetric CDW to the nematic $3Q$ phase and eventually to a $1Q$ stripe state with increasing electron doping. In Section~\ref{sec:TB}, we construct an effective 2D lattice model for TiSe$_2$ and reproduce the previous findings within extensive self-consistent mean-field calculations. In Section~\ref{sec:bulk} we extend our $\boldsymbol{k}\cdot\boldsymbol{p}$ analysis to 3D, obtaining the same sequence of transitions to $C_3$-breaking CDW states with electron doping. Finally, in Section~\ref{sec:discussion} we summarize our findings and discuss their signatures in the context of past experiments as well as new proposals.}

{\section{Phenomenology of the CDW transition in TiSe$_2$\label{sec:phenomenology}}} 

TiSe$_2$ is a layered dichacogenide that crystalizes in the 1T structure made of Se-Ti-Se triangular layers arranged in ABC stacking. The high temperature space group is $P \bar{3}m1$ (\#164, point group $D_{3d})$. Neutron diffraction \cite{DMW76,Wakabayashi78} established the emergence of a superlattice with CDW wavevectors $\vec Q = \vec L_i$ with atomic displacements transverse to $\vec Q$. Assuming the condensation of a single phonon this requires an  $L_1^-$ or $L_2^+$ irrep, and the structure refinement selected $L_1^-$. This finding was confirmed by X-ray scattering which observed the softening to zero energy of the $L_1^-$ phonon at the CDW transition \cite{Holt01}. Ref. \cite{DMW76} also determined that the ground state has order parameter $\vec \Delta = (\Delta,\Delta,\Delta)$ which leads to the space group $P\bar{3}c1$ (\#165, point group $D_{3d})$. The $L_1^-$ phonon displacement pattern leads to a trimerization of TiSe$_2$ units, as shown in Fig.~\ref{FigPhonons}(a,b,c). As an $L$ point modulation, it shows opposite displacements in adjacent TiSe$_2$ layers, leading to an interlayer inversion center that makes the structure centrosymmetric for any configuration of $\vec \Delta$. The CDW phase transition is preserved for any thickness from bulk to monolayer limit, with a small increase of $T_c$ as thickness is decreased \cite{CCF15,Chen16,Fang17}. The monolayer order parameter has symmetry $M_1^-$ (in the bulk, $M_1^-$ represents the same intralayer displacement as $L_1^-$, but the modulation in adjacent layers is in phase). 

\subsection{The chiral CDW proposal}

The threefold symmetric state in the bulk \cite{DMW76} is incompatible with several STM experiments~\cite{ILS10,Ishioka11,Iavarone12,Kim24seamless}, which clearly observed three CDW Bragg peaks $(I_1,I_2,I_3)$ with different intensities. This observation motivated the proposal of an alternative CDW distortion, known as the chiral CDW ~\cite{vanWezel11,van2012chiral,Zenker13,GW15,Peng22}, with space group C2 (SG \#5) which breaks $C_3$, all mirrors and inversion. The proposal heuristically introduced a complex order parameter for the transition $\Psi_i = |\Psi_i| e^{{-}i \varphi_i}$ and obtained the particular configuration $\varphi_1=0$ and $\varphi_2=-\varphi_3 \equiv \varphi$~\cite{vanWezel11,van2012chiral} which leads to a chiral state. It should be noted, however, that when the CDW wavevector is half of a reciprocal lattice vector, $\boldsymbol{Q}_i = \tfrac{1}{2} \boldsymbol{G}_i$, order parameters corresponding to irreducible representations (irreps) of the little group are real~\cite{Nagaosa84,Sahu85}. For any other wavevector, the order parameter is complex, encoding the independence of the modulations at $\boldsymbol{Q}_i$ and $-\boldsymbol{Q}_i$, as in the well-known cases of NbSe$_2$ ($\boldsymbol{Q}_i = \tfrac{1}{3} \boldsymbol{G}_i$), and the chiral structure of the elemental chalcogens Se and Te~\cite{Fukutome84,Silva18,Silva18SP} ($\boldsymbol{Q} =\tfrac{1}{3} (\boldsymbol{G}_1 + \boldsymbol{G}_2 + \boldsymbol{G}_3)$). For TiSe$_2$, however, the proposed complex order parameter is not a proper irrep of the space group, but rather the combination of two irreps. This can be seen by decomposing the phonon displacements
\begin{align}
    &\boldsymbol{u} (\boldsymbol{x}) = {\rm Re}  \sum_i \boldsymbol{u}_i \Psi_i e^{i (\boldsymbol{Q}_i \cdot \boldsymbol{x})} \nonumber \\= &\sum_i  \boldsymbol{u}_i |\Psi_i| \cos(\varphi_i) \cos(\boldsymbol{Q}_i \cdot \boldsymbol{x}) + \boldsymbol{u}_i |\Psi_i| \sin(\varphi_i) \sin(\boldsymbol{Q}_i \cdot \boldsymbol{x})
\end{align}
with fixed in-plane polarization vector $\boldsymbol{u}_i$ transverse to $\boldsymbol{Q}_i$. The order parameter $\Delta_i = |\Psi_i| \cos \varphi_i$ gives the standard displacement with symmetry $L_1^-$,  while the order parameter $\Phi_i = |\Psi_i| \sin \varphi_i$ gives displacements with symmetry $L_2^+$ \cite{Peng22}, plotted here in Fig.~\ref{FigPhonons}(d,e). The chiral CDW proposal, represented in Fig.~\ref{FigPhonons}(f), is therefore a mixture of two independent phonons with different symmetries, an $L_1^-$ phonon in a nematic $3Q$ state $\Delta_i=[1,\cos(\varphi),\cos(\varphi)]$ and an $L_2^+$ phonon in a $2Q$ state $\Phi_i= [0,\sin(\varphi),-\sin(\varphi)]$. The $L_1^-$ nematic $3Q$ state alone would have space group C2/c, preserving both $C_{2x}$ and inversion, but the addition of the extra $L_2^-$ phonon in the chiral CDW breaks the inversion symmetry, reducing the space group to C2. 

Given this, it is worth considering whether the nematic $3Q$ ground state of the $L_1^-$  phonon originally established in X-rays (or $M_1^-$ for monolayers) is actually compatible with the STM experiments. Since STM is a surface experiment, it can only probe the breaking of surface symmetries. The surface group of TiSe$_2$ is $C_{3v}$ and only contains threefold rotations $C_3$ and three vertical mirrors $\sigma_v$. Thus STM cannot probe the breaking of inversion $I$, a twofold axis $C_{2x}$ or a glide $M_x\{0,0,\tfrac{1}{2}\}$. The symmetry groups of the possible ground states for both $L_1^-$ and $M_1^-$ phonons are shown in Table~\ref{tab:symmetry_groups_monolayer}. 

Since three finite CDW Bragg peaks are observed, and we want $C_3$ breaking states, we may consider $(\Delta_1,\Delta_2,\Delta_2)$ or $(\Delta_1,\Delta_2,\Delta_3)$. Since the state $(\Delta_1,\Delta_2,\Delta_2)$ only preserves $C_{2x}$ (for $M_1^-$) or $C_{2x}$, $I$ and $M_x\{0,0,\tfrac{1}{2}\}$ (for $L_1^-$), it actually breaks \emph{all surface symmetries} and generically leads to three surface CDW Bragg peaks $I_1 \neq I_2 \neq I_3$. Therefore both $(\Delta_1,\Delta_2,\Delta_2)$ or $(\Delta_1,\Delta_2,\Delta_3)$ would be compatible with the STM observation. This can also be intuitively seen for \textbf{$(\Delta_1,\Delta_2,\Delta_2)$} from Fig. \ref{FigPhonons}(c), where it can be observed that the four topmost Se atoms are generically inequivalent. In addition, the state $(0,\Delta,\Delta)$ only preserves a surface mirror, which generically leads to $I_1\neq I_2=I_3$ (with $I_1$ allowed to second order in $\Delta$). For completeness, the chiral CDW state, which only preserves a single $C_{2x}$ axis that is not a surface symmetry, would again be compatible with $I_1 \neq I_2 \neq I_3$. STM alone cannot therefore distinguish between the nematic $L_1^-$ $3Q$ CDW and the chiral CDW states.

\begin{table*}[t]
\caption{Symmetry groups of the possible ordered phases in TiSe$_2$ (see also Ref. \cite{Subedi22}). The first three columns refer to an $M_1^-$ order parameter, relevant for the monolayer case, while the next three refer to an $L_1^-$ order parameter, relevant for the bulk case. Note the state $\vec{\Delta} = (\Delta_1,\Delta_2,\Delta_3)$ is included for completeness but never found as a ground state in this work.}
\label{tab:symmetry_groups_monolayer}
\begin{tabularx}{1\linewidth}{LLL|LLL}
\hline \hline
$\vec{\Delta} \; (M_1^-)$                 & SG             & PG & $\vec{\Delta} \;(L_1^-)$                 & SG             & PG       \\ \hline
$(\Delta,\Delta,\Delta)$       & P321 ($\#149$) & $D_3$ & $(\Delta,\Delta,\Delta)$       & $P \bar{3}c1$ ($\#165$) & $D_{3d}$    \\ 
$(\Delta,0,0)$                 & {P2/c ($\#13$)}  & $C_{2h}$ &$(\Delta,0,0)$                 & C2/c ($\#15$)  & $C_{2h}$ \\
$(0,\Delta,\Delta)$            & C2/m ($\#12$)  & $C_{2h}$& $(0,\Delta,\Delta)$            & C2/m ($\#12$)  & $C_{2h}$ \\
$(\Delta_1,\Delta_2,\Delta_2)$ & C2 ($\#5$)     & $C_{2}$& $(\Delta_1,\Delta_2,\Delta_2)$ & C2/c ($\#15$)     & $C_{2h}$  \\ 
$(\Delta_1,\Delta_2,\Delta_3)$ & P1 ($\#1$)     & 1 & $(\Delta_1,\Delta_2,\Delta_3)$ & P$\bar{1}$ ($\#2$)     & $C_i$        \\ \hline \hline
\end{tabularx}
\end{table*}

\subsection{Experimental survey of further symmetry breaking}
\label{subsec:intro_sym_break}

Since STM evidence is compatible with both the chiral CDW and the nematic $3Q$ state of $L_1^-$, it is important to consider whether other experimental bulk probes have provided evidence for the chiral CDW, in particular by detecting the breaking of inversion symmetry. X-ray and thermodynamic evidence for a second chiral transition were reported soon after the proposal~\cite{Castellan13}, but the X-ray evidence was later contested~\cite{Lin19,Ueda21} and explained in terms of the standard $L_1^-$ CDW. The thermodynamic evidence for a second transition remains, but its nature has not been established. Resonant inelastic X-Ray scattering is a useful tool to probe electronic ordering, and Ref. \cite{Peng22} emphasized the existence of orbital order in TiSe$_2$, but not necessarily inversion breaking. More recently, X-ray circular dichroism was reported as a signature of the chiral CDW \cite{Xiao24} but this experiment has also been interpreted in terms of the standard Di Salvo state \cite{Ueda25} and it is controversial whether it represents direct evidence of inversion symmetry breaking. 

The breaking of inversion symmetry can also be probed with nonlinear optical effects like second harmonic generation (SHG) and photogalvanic effects (PGE), which are activated below the transition temperature to a non-centrosymmetric structure.  The SHG in TiSe$_2$ has a weak temperature dependence as $T_{\rm CDW}$ is crossed, it is also observed in TiS$_2$ without a CDW transition, and it decays with the number of layers \cite{ZhangSHG22}, which suggests that the SHG is just a surface effect or arises from a higher order (quadrupole) contribution which does not require inversion breaking. Odd harmonic generation, which does not require broken inversion, does show an abrupt activation at the phase transition, and displays signatures of rotational anisotropy \cite{Tyulnev24}. Finally, the circular photogalvanic effect does not activate at $T_{\rm CDW}$ under standard conditions, but only when samples cooled in the presence of circularly polarized light ~\cite{Xu20,jog_optically_2023}. This suggests that light can be used to change the CDW into a metastable state which breaks inversion symmetry, but that the true ground state is inversion preserving~\cite{Wickramaratne22}. 

Finally, the comparison of Raman and infrared spectra is often used to detect inversion symmetry breaking. Since in the presence of inversion, only odd-parity modes are seen in infrared while only even modes are seen in Raman, the presence of modes at the same energy in both probes is often taken as a signature that inversion is broken. In TiSe$_2$, this analysis is complex due to the presence of many phonons~\cite{Uchida81,Holy77,Barath08,hellgren_critical_2017}, but a recent Raman experiment has claimed the breaking of both inversion and $C_3$ symmetry \cite{kim2024origin}, although the X-ray refinement is considered consistent with the Di Salvo state.

\subsection{The conduction band population}

In this work, we argue that the observed broken symmetries in the CDW state of TiSe$_2$ are very sensitive to the conduction band population, which has varied across the literature in different experiments, often unintentionally. TiSe$_2$ samples are usually synthesized using the chemical vapor transport (CVT) method. While stoichiometric samples should either be insulating or compensated semimetals, CVT-grown samples exhibit a certain amount of Se deficiency, which effectively electron dopes the system. This is reflected in the low temperature metallic behavior of CVT-grown TiSe$_2$ within the CDW phase~\cite{Campbell19,WBK19}. The observation of the bottom of the conduction band in ARPES further proves this scenario~\cite{Watson19,Watson20,Campbell19}. Ref.~\cite{Campbell19} used a different growth method at high pressure using argon gas, and for the first time observed no Se deficiency, insulating low-temperature transport, a reduced carrier concentration in Hall measurements, and no occupied conduction band in ARPES. Thus, these samples are truly insulating, while all the previous ones are metallic due to the electron doping. Remarkably, all the samples with signatures for further symmetry breaking within the CDW were CVT-grown, using either Se flux~\cite{ILS10,Ishioka11} or iodine~\cite{Peng22,Iavarone12,Castellan13}
as transport agents, and therefore contained a certain amount of electron doping (Refs.~\cite{Iavarone12,Novello17} have also demonstrated the breaking of $C_3$ symmetry in STM measurements in Cu-doped samples). The precise amount of carriers in each of the experiments is nevertheless much harder to establish. 

In this paper, we explore the possible connection between the electron doping and the breaking of further symmetries within the CDW, and we find that nematic and stripe CDW states can indeed be induced by doping. The uncontrolled amount of electron doping in different samples might therefore explain the contradicting experimental claims regarding the symmetry of the CDW.

\section{Continuum $\mathbf{k\cdot p}$ model}
\label{sec:kp}

\subsection{Continuum model for the unreconstructed bands}

We begin by constructing a continuum $\boldsymbol{k} \cdot \boldsymbol{p}$ model for the valence and conduction bands of TiSe$_2$, constrained by the symmetries of the high temperature phase (space group $P\bar{3}m1$, point group $D_{3d}$). The band structure near the Fermi level consists of three electron pockets at the $L$ points with $L_1^+$ symmetry, which derive from Ti $d$ orbitals, and two hole pockets of $\Gamma_3^-$ symmetry located at $\Gamma$ \cite{zunger_band_1978}, which derive from Se $p$ orbitals, as shown in Fig.~\ref{Fig1}(c,d). The presence of a small indirect gap ($E_g>0$) or overlap ($E_g<0$) is still debated \cite{rasch_semi_2008,mottas_semimetal--semiconductor_2019,Watson19,jaouen_phase_2019,Watson20}, but our conclusions apply to both cases. The CDW order parameter has $L_1^-$ symmetry and hybridizes the valence and conduction bands. In the case of monolayer TiSe$_2$, the electron pockets occur at the $M$ points and have $M_1^+$ symmetry, resulting in an order parameter with $M_1^-$ symmetry. Our proposed mechanism for rotation symmetry breaking applies to both monolayer and bulk samples with the corresponding replacement of the symmetry labels, so for simplicity we now focus on the monolayer {(see Sec.~\ref{sec:3D_kp_model} for the 3D model)}.

The Hamiltonian $H_{dd}^0$ of the three anisotropic electron pockets, expressed in the basis $\{d_1,d_2,d_3\}$, where $d_i$ represents the band at $M_i$, is:
{
\begin{align} 
    H_{dd}^0(\boldsymbol{k}) = \begin{pmatrix}
        \varepsilon_{d,1}(\boldsymbol{k}) & 0 & 0 \\
        0 & \varepsilon_{d,2}(\boldsymbol{k}) & 0 \\
        0 & 0 & \varepsilon_{d,3}(\boldsymbol{k}) \\
    \end{pmatrix} 
\label{eq:H_dd}
\end{align}
with 
\begin{align}
    & \varepsilon_{d,1}(\boldsymbol{k}) = a_d (k_x^2+k_y^2) + b_d(k_x^2-k_y^2), \label{eq:Ed1} \\
    & \varepsilon_{d,2}(\boldsymbol{k}) = a_d (k_x^2+k_y^2) + b_d[-\sqrt{3}k_xk_y -\tfrac{1}{2}(k_x^2-k_y^2)], \\
    & \varepsilon_{d,3}(\boldsymbol{k}) = a_d (k_x^2+k_y^2) + b_d[\sqrt{3}k_xk_y -\tfrac{1}{2}(k_x^2-k_y^2)],\label{eq:Ed3}
\end{align}}
{where $\boldsymbol{k} = (k_x,k_y)$ (see Apps.~\ref{app:group_theory} and \ref{app:kp_construction} for details on the symmetry analysis and derivation of the model)}. The ellipticity of the electron pockets is parametrized by $b_d$, and the fact that the long axis is along $\Gamma M$ requires $b_d>0$. \textit{Ab initio} calculations predict a large value of the ellipticity, $b_d/a_d \sim 0.87$ \cite{monney_impact_2015,chen_reproduction_2018,guster_first_2018}, which we take as the realistic value for the rest of this work. This value is also within the range of the experimentally reported conduction band masses (see Table~\ref{table:masses_fit} in App.~\ref{app:TB_params}).

The hole bands at $\Gamma$ are represented by the Hamiltonian $H_{pp}^0$, which in the $\{p_x,p_y\}$ basis reads
\begin{align}
    H_{pp}^0(\boldsymbol{k}) = \left(\begin{array}{cc}
    a_p k^2  + b_p(k_x^2-k_y^2) & b_p 2k_xk_y \\
    b_p 2k_xk_y & a_p k^2 -b_p(k_x^2-k_y^2)
    \end{array}
    \right),
\label{eq:Hpp0}
\end{align}
where $b_p$ parametrizes the orbital texture of the hole bands.  \textit{Ab initio} calculations~\cite{zunger_band_1978} show the top valence band has $p_x$ character along the $\Gamma M$ line (it is mirror $M_x$ odd, becoming $M_1^-$ at $M$), which sets $b_p<0$. We take the value $b_p/a_p= 0.25$ from Refs.~\cite{kolekar_controlling_2018,guster_first_2018}.

\begin{figure}
    \centering
    \includegraphics[width=0.48\textwidth]{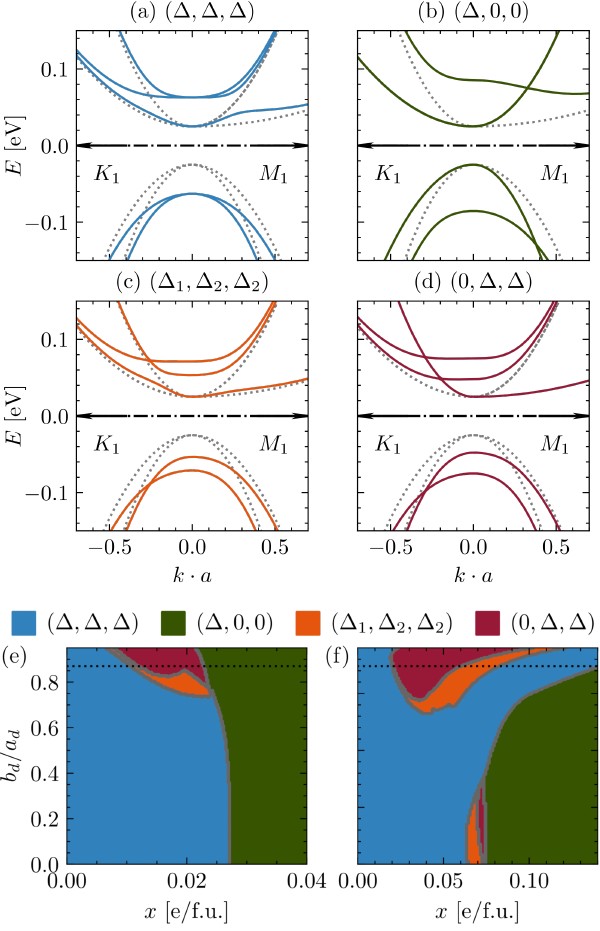}
    \caption{(a,b) Folded bands in the presence of the order parameter $\vec{\Delta}$ for directions (a) $(\Delta,\Delta,\Delta)$, (b) $(\Delta,0,0)$, (c) $(\Delta_1,\Delta_2,\Delta_2)$ with $\Delta_1=\tfrac{1}{2}\Delta_2$, and (d) $(0,\Delta,\Delta)$. Grey dashed lines show the bands for $|\vec{\Delta}|=0$. 
    (e) Constant $|\vec{\Delta}|$ phase diagram for $E_g=+50\mathrm{meV}$ and $|\vec{\Delta}| = {82} \mathrm{meV}$ as a function of electron doping $x$ and ellipticity of the conduction bands $b_d/a_d$. The black dotted line indicates the realistic ellipticity $b_d/a_d = 0.87$. 
    (f) The same as (e) but with $E_g = -100\mathrm{meV}$ and on a larger $x$ range.}
    \label{Fig2}
\end{figure}

Anticipating the CDW phase transition, we work in the folded BZ where the $M$ point is folded to $\Gamma$. In the basis $\{d_1,d_2,d_3,p_x,p_y\}$, the bare $\boldsymbol{k} \cdot \boldsymbol{p}$ Hamiltonian is therefore
\begin{align}
    H^0(\boldsymbol{k}) = \left(\begin{array}{cc}
    \frac{E_g}{2} + H_{dd}^0(\boldsymbol{k}) & 0 \\
    0 & -\frac{E_g}{2} + H_{pp}^0(\boldsymbol{k})
    \end{array}
    \right),
\label{eq:H0_kp}
\end{align}
where $E_g$ is the gap, which is at most $|E_g|< 100\mathrm{meV}$. We do not consider spin-orbit coupling, because it does not significantly affect the CDW~\cite{hellgren_critical_2017}. The bands of the model~\eqref{eq:H0_kp} are shown as grey dotted lines in Fig.~\ref{Fig2}(a,b{,c}).

\subsection{Continuum model in the presence of the CDW}

{Next, we consider the effect of the CDW with $M_1^-$ symmetry at the mean-field level, which is characterized by an order parameter $\vec{\Delta} = (\Delta_1,\Delta_2,\Delta_3)$ that couples the conduction and valence bands.} This order parameter can be thought of as the $M_1^-$ phonon displacement in normal coordinates, or as the excitonic order parameter of the same symmetry obtained from a mean-field decoupling. {Both excitonic and phononic contributions to the order parameter enter the mean-field $\boldsymbol{k}\cdot\boldsymbol{p}$ electronic Hamiltonian in the same way, which only depends on the $M_1^-$ symmetry of the CDW. In particular, the total electronic energy determined using the $\boldsymbol{k}\cdot\boldsymbol{p}$ model is agnostic to the microscopic origin of the order parameter $\vec{\Delta}$. The total 2D $\boldsymbol{k}\cdot\boldsymbol{p}$  Hamiltonian in the presence of the $M_1^-$ CDW becomes}
\begin{equation}
    H = \left(\begin{array}{cc}
    \frac{E_g}{2} + H_{dd}^0(\boldsymbol{k}) & H_{dp}(\vec{\Delta}) \\
    H_{dp}^\dagger(\vec{\Delta}) & -\frac{E_g}{2} + H_{pp}^0(\boldsymbol{k})
    \end{array} \right).
    \label{eq:H_kp}
\end{equation}
{As derived in App.~\ref{app:kp_construction}, to lowest-order in $|\vec{\Delta}|$ and $\boldsymbol{k}$, the symmetry-allowed coupling to the $M_1^-$ order parameter $\vec{\Delta}$ is $\Delta_1 d\dag_1 p_x + \Delta_2 d\dag_2 \left( -\tfrac{1}{2} p_x + \tfrac{\sqrt{3}}{2} p_y \right) + \Delta_3 d\dag_3 \left( -\tfrac{1}{2} p_x - \tfrac{\sqrt{3}}{2} p_y \right) + \mathrm{h.c.}$, so that the interband term in the Hamiltonian becomes} 
\begin{align}
    H_{dp}^\dagger(\vec{\Delta}) = \left(\begin{array}{ccc}
    \Delta_1 &  -\tfrac{1}{2}\Delta_2 & -\tfrac{1}{2}\Delta_3 \\
    0 & \tfrac{\sqrt{3}}{2}\Delta_2 &  -\tfrac{\sqrt{3}}{2}\Delta_3 \\
    \end{array}
    \right).
\label{eq:HDelta}
\end{align}

The $C_3$-symmetric $3Q$ state \cite{DMW76} is represented by $\vec{\Delta} = \tfrac{|\vec{\Delta}|}{\sqrt{3}}(1,1,1)$. This order parameter causes a repulsion between the doublet of valence bands and a doublet of conduction bands (see Fig.~\ref{Fig2}(a)), leaving the band edge of the third conduction band unaffected by the CDW transition, as seen in ARPES~\cite{Watson19,Watson20}. Figs.~\ref{Fig2}(b,c,d) show the bands for the $1Q$ $\vec{\Delta} = |\vec{\Delta}|(1,0,0)$, {nematic $3Q$ $\vec{\Delta} = (\Delta_1,\Delta_2,\Delta_2)$,} and $2Q$ $\vec{\Delta} = \tfrac{|\vec{\Delta}|}{\sqrt{2}}(0,1,1)$ states, respectively. In the $1Q$ case, only one conduction band is repelled to high energies.

\subsection{Energetics of the CDW order parameter}

The CDW order parameter has three symmetry-related components, allowing for distinct ground states. {This raises the questions of whether the $C_3$-symmetric $3Q$ state is the ground state, and whether it is stable upon doping.} These questions can be addressed within the continuum $\boldsymbol{k} \cdot \boldsymbol{p}$ model without knowledge of the precise nature or structure of the interaction that gives rise to $\vec{\Delta}$. Assuming an interaction that only depends on $|\vec{\Delta}|^2 = \Delta_1^2 + \Delta_2^2+\Delta_3^2$, which holds when any local four-fermion interaction is decoupled only in the chosen channel, the direction of $\vec{\Delta}$ for fixed magnitude can be obtained at zero temperature by minimizing the energy of the occupied bands $E = \int d^2k/(2\pi)^2 \sum_n \varepsilon_n(\boldsymbol{k}) \theta(\mu-\varepsilon_n(\boldsymbol{k}))$ where $H \psi_n(\boldsymbol{k}) = \varepsilon_n(\boldsymbol{k}) \psi_n(\boldsymbol{k})$. A phase diagram with approximate phase boundaries (compared to a self-consistent calculation of $\vec{\Delta}$) can thus be obtained to explain the origin of the different phases.

In Fig.~\ref{Fig2}(e) we show such a phase diagram for fixed $|\vec{\Delta}|$ and $E_g>0$, as a function of doping $x$ and conduction band ellipticity $b_d/a_d$. We vary $a_d$ to keep the bare density of states (DOS) $\rho=1/(4\pi\sqrt{a_d^2-b_d^2})$ constant, to emphasize the role of ellipticity. {Four} main phases are observed. The $C_3$-symmetric $3Q$ state is the ground state at stoichiometry for any $b_d$, as expected. For moderate values of $b_d/a_d < 0.7$, there is a sharp $3Q$ to $1Q$ transition at a critical doping $x_{1Q}$. In addition, at higher values of $b_d/a_d$, a nematic $3Q$ phase emerges where $\vec{\Delta} = (\Delta_1,\Delta_2,\Delta_2)$, with $|\Delta_1| \neq |\Delta_2|$. Further increasing $b_d/a_d$, $\Delta_1$ vanishes and the nematic phase becomes $2Q$ with $\vec{\Delta} = (0,\Delta,\Delta)$. This $2Q$ state is distinguished from the nematic $3Q$  because it preserves an extra inversion symmetry in the center of the Ti-Ti bond {(see Table~\ref{tab:symmetry_groups_monolayer}). We note that, although we have considered the $(\Delta_1, \Delta_2, \Delta_3)$ state in our calculations, it does not appear as the lowest-energy state for the parameter space we have explored within the numerical precision.}

The emergence of the $3Q$ to $1Q$ transition can be understood most clearly in the isotropic case $b_p=b_d=0$, for which the energies of the two phases can be computed and integrated analytically (see App.~\ref{app:kp_energy_analytic}). This yields a critical doping which takes the simple form $n_{1Q} = \tfrac{|\vec{\Delta}|}{\pi} \sqrt{\tfrac{\log 2}{{2}a_d(a_d-a_p)}}$ in the $E_g=0$ limit. To see that a transition from the $3Q$ to the $1Q$ state should occur at some critical doping, consider the effect of adding carriers on the energies of these states. In the $1Q$ state two bands are populated as carriers are added, whereas in the $3Q$ state only one band is populated by additional carriers, as can be seen in Fig.~\ref{Fig2}(a,b). This implies a higher chemical potential for the $3Q$ state and an energy which increases faster relative to the $1Q$ state, eventually making the latter lower in energy. This mechanism is still at work at finite $b_d$, as we observe numerically {in Fig.~\ref{Fig2}(e)}.

\begin{figure}
    \centering
    \includegraphics[width=0.48\textwidth]{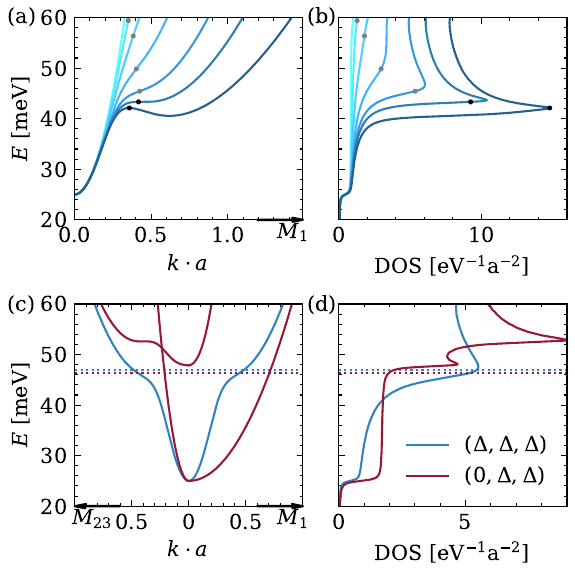}
    \caption{(a) Lowest conduction band along the $\Gamma M$ direction and corresponding DOS in the $C_3$-symmetric $3Q$ phase with $E_g=+50 \mathrm{meV}$ and $|\vec{\Delta}| = {82} \mathrm{meV}$ for constant $a_d$ and ellipticities $b_d/a_d = 0, 0.39, 0.60, 0.78, 0.88, 0.92, 0.95$. 
    A kink in the band is signaled by gray dots, and a van Hove singularity by black dots. (b) Corresponding DOS for each band in (a). (c) Conduction bands along the $\Gamma M_1$ and $\Gamma M_2$ directions ($\Gamma M_3$ is equivalent to $\Gamma M_2$) in the $C_3$-symmetric $3Q$ (blue) and nematic 
    $2Q$ (red) phases with $E_g=+50 \mathrm{meV}$, $|\vec{\Delta}| = {82}\mathrm{meV}$ and $b_d/a_d = 0.87$. Horizontal dotted lines indicate the chemical potentials for $x=0.0182$, where the $C_3$-symmetric $3Q$ phase is unstable towards the  
    $2Q$ state. (d) Corresponding DOS for each band in (c). The three equivalent vHs singularities of the lowest conduction band in the symmetric $3Q$ state disappear in the $2Q$ state, which has higher DOS at lower energy and therefore lower total energy.}
    \label{Fig3}
\end{figure}

To understand the origin of the $3Q/2Q$ nematic phase, Fig.~\ref{Fig3}(a) shows a close up of the dispersion of the lowest conduction band $\varepsilon_{c1}$ along $\Gamma M$ as a function of $b_d$ for fixed $a_d$ and $\vec{\Delta} = \frac{|\vec{\Delta}|}{\sqrt{3}} (1,1,1)$. Fig.~\ref{Fig3}(b) shows the corresponding DOS. Increasing the ellipticity first produces a kink in the dispersion (a relative minimum of $d\varepsilon_{c1}/dk_y$), which eventually gives rise to a van Hove singularity (vHs) with diverging DOS. When the filling is close to that of the six symmetry equivalent vHs points, breaking the $C_3$ symmetry can lower the energy by splitting these saddle points in energy, a known mechanism for nematicity in the doped honeycomb lattice \cite{Valenzuela08} and in Kagome superconductors \cite{Kiesel13}. Indeed, such splitting is observed in Fig.~\ref{Fig3}(c,d) in the nematic state, explaining its origin. In Fig.~\ref{Fig2}(d), this $3Q/2Q$ nematic phase develops even at values of $b_d$ where the vHs singularity is not fully developed and there is only a finite but sizable DOS peak. Eventually, for higher doping, the $1Q$ stripe phase always develops.

Fig.~\ref{Fig2}(f) shows the phase diagram for $E_g = -100 \mathrm{meV}$, which exhibits the following differences with respect to $E_g>0$. First, there is an intermediate nematic $3Q/2Q$ state also at small ellipticity, which occurs at fillings where the second conduction band, of approximate mexican hat shape for $\vec{\Delta} = \frac{|\vec{\Delta}|}{\sqrt{3}} (1,1,1)$, begins to be populated. Since the DOS is also large there, a similar mechanism as the one for large ellipticity drives the transition to the nematic $3Q/2Q$ phase. Furthermore, at high ellipticity, including $b_d/a_d\sim0.87$, a reentrant $C_3$-symmetric $3Q$ phase appears between the nematic $3Q$ state and the $1Q$ stripe phase. In this region, the doping is well above the vHs, so that no nematic instability occurs, but the $1Q$ state energy is still higher. This reentrant phase shows that the mechanisms for the $1Q$ and $2Q/3Q$ nematic states are generically different. 
\\

\section{Minimal lattice model}
\label{sec:TB}

\subsection{Model}

The continuum $\boldsymbol{k} \cdot \boldsymbol{p}$ model provides a compelling basic understanding of the CDW phase diagram as a function of doping, purely based on energetic considerations. To obtain a more refined understanding based on a model which includes microscopic interactions, we now consider an effective tight-binding model and study its ground state phase diagram within Hartree-Fock theory. {The realistic tight-binding model of TiSe$_2$ would consist of at least 7 orbitals per unit cell: the four $\{p_x,p_y\}$ orbitals from the two Se atoms, and the approximate $t_{2g}$ triplet of $d$ orbitals $\{d_{ab},d_{bc},d_{ca}\}$ from the Ti, where $a,b,c$ are the approximately cubic axes pointing along the Ti-Se bonds (see e.g. Ref.~\cite{Kaneko18} for an 11-orbital model). To simplify the problem, here we} construct a minimal {three-orbital} tight-binding model which accurately captures the band dispersion and eigenstate symmetry near the Fermi level and reproduces the low-energy $\boldsymbol{k} \cdot \boldsymbol{p}$ model when expanded near $\Gamma$ and $M$. As such, the tight-binding model can be viewed as a lattice regularization of the $\boldsymbol{k} \cdot \boldsymbol{p}$ model. Furthermore, the tight-binding model breaks the artificial independent conservation of the charge of the conduction and valence bands present in the $\boldsymbol{k}\cdot\boldsymbol{p}$ model.

\begin{figure}[!t]
    \centering
    \includegraphics[width=0.48\textwidth]{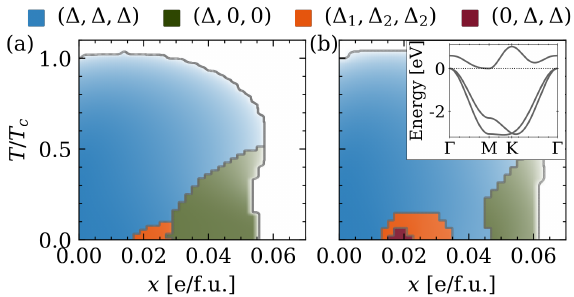}
    \caption{
    Temperature-doping phase diagrams for positive and negative gaps: 
    (a) $E_g = +25 \mathrm{meV}$, $V_{dp} = 845 \mathrm{meV}$; 
    (b) $E_g = -35 \mathrm{meV}$, $V_{dp} = 716.5 \mathrm{meV}$. 
    The intensity of the color of each phase is proportional to $|\vec{\Delta}|$. 
    Critical temperatures at charge neutrality are (a) $T_c = 459 \mathrm{K}$ and (b) $T_c = 239 \mathrm{K}$.
    The inset shows the band structure of the tight-binding model in the unfolded unit cell for $E_g=0$. }
    \label{Fig4}
\end{figure}

Our minimal three-orbital lattice model exhibits $d_{z^2},p_x,p_y$-like orbitals sitting at the same 2D triangular lattice sites. The $d_{z^2}$ orbital transforms as $A_{1g}$ and constitutes the conduction band, whereas the $\{p_x,p_y\}$ orbitals transform as $E_u$ and form the valence bands. We include hoppings up to third nearest neighbours for $d_{z^2}$, nearest neighbour $\sigma$ and $\pi$ hopping for $\{p_x,p_y\}$, and an interorbital hopping $t_{dp}$:
\begin{equation}
\begin{split}
    & H_0 = \sum_i \left( \varepsilon_d d_i^\dagger  d_i + \varepsilon_p \boldsymbol{p}_i^\dagger \cdot \boldsymbol{p}_i \right) + \sum_{n=1}^3 \sum_{\langle i j \rangle_n} t_{dd}^{(n)} d_i^\dagger  d_j \\ 
    & - \hspace{-3pt} \sum_{\langle i j \rangle_1} \hspace{-3pt} \left[ \left( t_{pp\sigma} + t_{pp\pi} \right) (\boldsymbol{p}_i^\dagger \cdot \hat{\boldsymbol{r}}_{ij}) (\hat{\boldsymbol{r}}_{ij} \cdot \boldsymbol{p}_j) - t_{pp\pi} \boldsymbol{p}_i^\dagger \cdot \boldsymbol{p}_j \right] \\
    & - \hspace{-3pt} \sum_{\langle i j \rangle_1} i t_{dp} \left( d_i^\dagger \hat{\boldsymbol{r}}_{ij} \cdot \boldsymbol{p}_{j} - h.c. \right).
\end{split}
\label{eq:H0_TB}
\end{equation}
Here $\langle i j \rangle_n$ are the $n^{th}$ nearest neighbours, $\boldsymbol{p}_i = \left( p_{xi}, p_{yi} \right)$, and $\hat{\boldsymbol{r}}_{ij} = (\boldsymbol{r}_j-\boldsymbol{r}_i)/|\boldsymbol{r}_j-\boldsymbol{r}_i|$ is the unit vector from site $i$ to site $j$. As described in App.~\ref{app:TB_params}, the hopping parameters are fit to reproduce the gap and masses of the bands near the Fermi level (see the inset of Fig.~\ref{Fig4} for the bands for $E_g=0$).

The CDW order in this model is represented by the local $p$-$d$ hybridization, encoded in the fermion bilinear $\langle d_j^\dagger p_{\alpha j} \rangle$, with $j=1,...,4$ running over the sites of the 2x2 supercell, which transforms under symmetry as $\Gamma_3^- \oplus M_1^- \oplus M_2^-$ (see App.~\ref{app:MF_order_parameters}). We now complete the model with the most local interaction that is attractive only for the $M_1^-$ channel, which is {a nearest-neighbor inter-orbital dipole-dipole like interaction:}
\begin{equation}
    H_{\rm int} = V_{dp} \sum_{\langle i j \rangle_1} \left( d_i^\dagger \boldsymbol{p}_i \right)^\dagger \cdot \left( 1 - 2 \hat{\boldsymbol{r}}_{ij} \otimes \hat{\boldsymbol{r}}_{ij} \right) \cdot \left( d_j^\dagger \boldsymbol{p}_j \right),
\label{eq:Hint_TB}
\end{equation}
with $V_{dp}>0$. {The microscopic origin of this effective interaction can be thought of as the static limit of the projected electron-electron and electron-phonon interactions after phonons have been integrated out. This suggests that, while the quantitative value of the critical temperature is likely not captured by the effective tight-binding model, the symmetry of the CDW can be reproduced.}

{As derived in App.~\ref{app:MF_order_parameters}}, when decoupled in the mean-field $\langle d_j^\dagger p_{\alpha j} \rangle$ channel, the effective $V_{dp}$ interaction is only attractive for the $M_1^-$ channel, with an attraction $-2V_{dp}$. The time-reversal even $M_1^-$ component of $-2V_{dp}\langle d_j^\dagger p_{\alpha j} \rangle$ defines the order parameter $\vec \Delta$
\begin{align}
    \Delta_a = - {\frac{1}{2}} V_{dp} \sum_{j} e^{i \boldsymbol{M}_a \cdot \boldsymbol{r}_j} \mathrm{Re} \langle d_j\dag \boldsymbol{p}_j \rangle \wedge \frac{\boldsymbol{M}_a}{|\boldsymbol{M}_a|},
\label{eq:Delta}
\end{align}
where $a=1,2,3$ and $\boldsymbol{v} \wedge \boldsymbol{w} = v_x w_y - v_y w_x$. The constant prefactor in Eq.~\eqref{eq:Delta} is chosen so that $\vec \Delta$ couples to electrons and holes as in Eq.~\eqref{eq:H_kp} at low energies (see App.~\ref{app:MF_order_parameters}). Our interacting TB Hamiltonian therefore serves as a minimal model to analyze the energetics of the $M_1^-$ CDW order.

\subsection{Hartree-Fock phase diagram}

We perform self-consistent mean-field calculations of the Hamiltonian $H_0 + H_{\rm int}$ decoupled in the $\langle d_j^\dagger p_{\alpha j} \rangle$ channels for different values of the initial gap. We choose $V_{dp}$ such that the critical doping for the disappearance of the commensurate CDW is $\sim 6\%$, as determined experimentally~\cite{wu_transport_2007,LOL15,KogarICDW,Watson20}.

Figs.~\ref{Fig4}(a,b) display the resulting phase $T$-$x$ diagrams for gaps $E_g = + 25 \mathrm{meV}, - 35 \mathrm{meV}$. States $(\Delta,\Delta,\Delta)$, $(\Delta_1,\Delta_2,\Delta_2)$, and $(\Delta,0,0)$ appear for both gaps, while a $2Q$ $(0,\Delta,\Delta)$ phase develops only in the negative gap case. At zero temperature, the $C_3$-symmetric $3Q$ phase $(\Delta,\Delta,\Delta)$ appears at low doping, while the $1Q$-stripe order only develops at high doping, with the nematic $3Q$/$2Q$ phase developing in between, in agreement with Fig.~\ref{Fig2}(d,e). When the gap is negative, the reentrant $C_3$-symmetric $3Q$ phase is also observed between the nematic $3Q$ and the $1Q$ phases. Starting from high temperature, the first transition is always to the $C_3$-symmetric $3Q$ phase. At high doping, the $1Q$ phase develops at a lower but comparable temperature. On the other hand, the nematic $3Q$/$2Q$ phase only appears at low temperature ($T_{\text{nematic}} \sim 0.2 \hspace{3pt} T_{\mathrm{CDW}}$), which is expected for a secondary instability of the $3Q$ CDW state.

{\section{Bulk TiSe$_2$}}
\label{sec:bulk}

\begin{figure}[!b]
    \centering
    \includegraphics[width=1\linewidth]{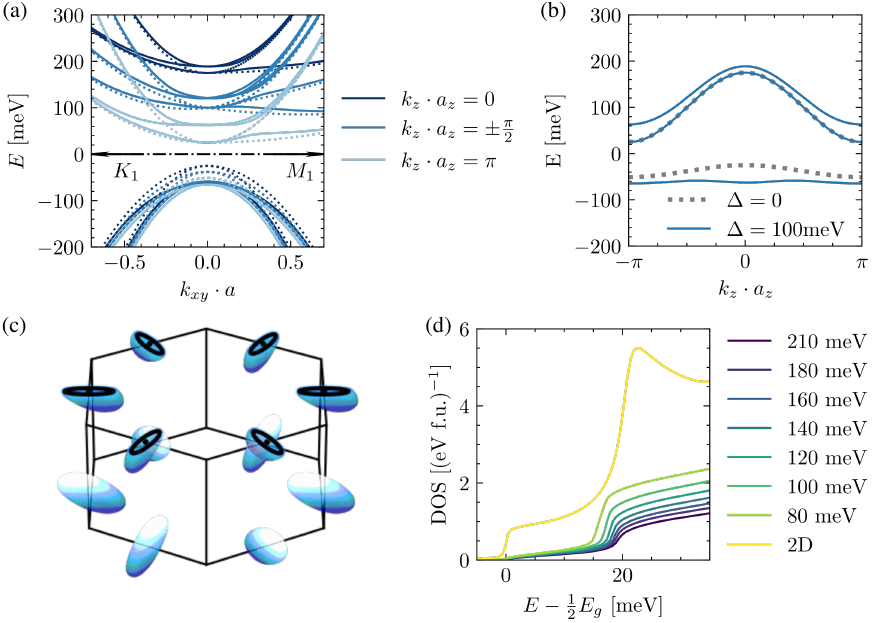}
    \caption{Band structure of the 3D $\boldsymbol{k}\cdot\boldsymbol{p}$ model of Eqs. (\ref{eq:H3D_ini}-\ref{eq:H3D_fin}) for $E_g=+50\mathrm{meV}$, $4 c_d = E(M) - E(L) = 150 \mathrm{meV}$ and $\theta_d=60\degree$. 
    (a) Bands as a function of the in-plane momentum $k_{xy}$ in the $K_1\Gamma M_1$ line for selected values of the out-of-plane momentum $k_z$. Dotted lines indicate the bands in the normal state, while full lines show the bands in the $C_3$-symmetric $3Q$ CDW phase with $|\vec{\Delta}|={82}\mathrm{meV}$.
    (b) $k_z$ dispersion of top of the valence band ($\Gamma A$) and the bottom of the conduction band ($M L$). 
    (c) Fermi surface of the normal state for a chemical potential $\mu=50\mathrm{meV}$.
    In (a-c) we have shifted the $k_z$ of the conduction band by $\pi$ so that $k_z=0$ corresponds to the $M$ point and $k_z=\pi$ corresponds to the $L$ point.
    (d) DOS per unit cell for varying $4 c_d = E(M) - E(L)$; the 2D DOS is also plotted for reference.}
    \label{fig:3D_bands}
\end{figure}

In this section, we generalize our $\mathbf{k}\cdot\mathbf{p}$ analysis to 3D with an $L_1^-$ order parameter, and we show that we also obtain doping-induced bulk transitions to nematic and stripe states.

\subsection{3D continuum $\mathbf{k}\cdot\mathbf{p}$ model}
\label{sec:3D_kp_model}

In bulk TiSe$_2$, the topmost valence bands are also the $\Gamma_3^-$ doublet at the $\Gamma$ point, while lowest-lying electron pockets occur at the $L$ points and have $L_1^+$ symmetry. This results in an order parameter with $L_1^-$ symmetry. The bulk $\boldsymbol{k}\cdot\boldsymbol{p}$ model follows from replacing the $M$ labels to $L$ in the monolayer model of Eq.~\eqref{eq:H_kp}, and adding the $k_z$ dispersion. Due to the weak van der Waals tunneling between layers, the out-of-plane dispersion is well described by the lowest-order trigonometric functions respecting the symmetries, which allows to reproduce the full $k_z$ dispersion in the Brillouin zone, from $k_z=-\tfrac{\pi}{a_z}$ to $k_z=\frac{\pi}{a_z}$, with $a_z$ the out-of-plane lattice constant. Having the full $k_z$ dispersion is useful e.g. for comparing the DOS per unit cell between the 2D and 3D cases. The resulting noninteracting $\boldsymbol{k}\cdot\boldsymbol{p}$ Hamiltonian capturing the full $k_z$ dispersion exhibits four new terms with respect to the 2D Hamiltonian $H^{0}$ of Eq.~\eqref{eq:H0_kp}, described by the coefficients $c_{p/d}$ and $d_{p/d}$:
\begin{widetext}
\begin{align}
    \varepsilon_{d1}^{(3\mathrm{D})}(\boldsymbol{k},k_z) & = \varepsilon_{d1}(\boldsymbol{k}) + 4 c_d \cos^2(\tfrac{1}{2} a_z k_z) + d_d \sin(a_z k_z) k_y, \label{eq:H3D_ini} \\
    \varepsilon_{d2}^{(3\mathrm{D})}(\boldsymbol{k},k_z) & = \varepsilon_{d2}(\boldsymbol{k}) + 4 c_d \cos^2(\tfrac{1}{2} a_z k_z) - d_d \sin(a_z k_z) \left(\frac{1}{2} k_y + \frac{\sqrt{3}}{2} k_x \right), \\
    \varepsilon_{d3}^{(3\mathrm{D})}(\boldsymbol{k},k_z) & = \varepsilon_{d3}(\boldsymbol{k}) + 4 c_d \cos^2(\tfrac{1}{2} a_z k_z) - d_d \sin(a_z k_z) \left(\frac{1}{2} k_y - \frac{\sqrt{3}}{2} k_x \right), \\
    H_{dd}^{(3\mathrm{D})}(\boldsymbol{k},k_z) & = {\rm diag}\left[\varepsilon_{d,1}^{(3\mathrm{D})}(\boldsymbol{k},k_z),\varepsilon_{d,2}^{(3\mathrm{D})}(\boldsymbol{k,k_z}),\varepsilon_{d,3}^{(3\mathrm{D})}(\boldsymbol{k},k_z)\right], \\
    H_{pp}^{(3\mathrm{D})}(\boldsymbol{k},k_z) & = H_{pp}^0(\boldsymbol{k}) + \left(\begin{array}{cc}
    - 4 c_p \sin^2(\tfrac{1}{2} a_z k_z) - d_p \sin(a_z k_z) k_y & -d_p \sin(a_z k_z) k_x \\
    -d_p \sin(a_z k_z) k_x & - 4 c_p \sin^2(\tfrac{1}{2} a_z k_z) + d_p \sin(a_z k_z) k_y
    \end{array}
    \right), \label{eq:H3D_fin}
\end{align}
\end{widetext}
where $\varepsilon_{di}(\boldsymbol{k},k_z)$ are defined in Eqs.~(\ref{eq:Ed1}-\ref{eq:Ed3}) and $H_{pp}^0(\boldsymbol{k},k_z)$ is given by Eq.~\eqref{eq:Hpp0}. We refer to App.~\ref{app:3D_kp_model} for further details on the construction of the 3D model. 

The $c_{p/d}$ terms, which reduce to the quadratic $k_z^2$ dispersion in the $z$ direction close to the high-symmetry points $\Gamma/A$ and $L/M$, respectively, correspond to the energy difference between the valence bands at $\Gamma$ and $A$ [$4 c_{p} = E_p(\Gamma) - E_p(A)$], and to the energy difference between the conduction bands at $M$ and $L$ [$4 c_{d} = E_d(M) - E_d(L)$]. From the ARPES experiment of Ref. \cite{Watson19}, $4 c_{p} = E_p(\Gamma) - E_p(A) \simeq 26 \mathrm{meV}$. Unfortunately, since the conduction bands are unoccupied, ARPES does not capture the $M$ pocket, and only restricts $4 c_{d} = E_d(M) - E_d(L) > 50 \mathrm{meV}$ in the normal state. DFT calculations predict $4 c_{d} = E_d(M) - E_d(L) \simeq 220 \mathrm{meV}$ \cite{Watson19,hellgren_critical_2017}. However, it should be noted that DFT calculations significantly overestimate the value of $E_p(\Gamma) - E_p(A)$ \cite{Watson19}, predicting it to be about $80 \mathrm{meV}$ \cite{Watson19,hellgren_critical_2017}, which is three times larger than found in ARPES. Therefore, we will leave $c_d$ as a tunable parameter in our calculations, and determine the ground state as a function of this parameter.

The $d_{p/d}$ terms couple the in-plane $k_{x,y}$ and out-of-plane $k_z$ dispersions. For a chemical potential intersecting the valence/conduction bands, $2\tfrac{(a_{p/d}-b_{p/d})}{d_{p/d}} = \tan(\theta_{p/d})$, where $\theta_{p/d}$ is the polar angle between the Fermi surface and the $z$ axis at $k_z=0$. For the conduction bands, we consider $\theta_{d}$ in the range $60\degree - 45\degree$, as extracted from the DFT Fermi surface shown in Ref. \cite{Watson19}, which leads to $d_{d} = 2\tfrac{(a_{d}-b_{d})}{\tan(\theta_{d})} = 50-87\mathrm{meV}$, where we have used the same values $a_d = 326 \mathrm{meV}$ and $b_d/a_d=0.87$ as in the 2D model. The valence bands display an angle $\theta_{p} \simeq 90\degree$, so we will set $d_p = 0$.

The $L_1^-$ CDW order parameter $\vec{\Delta}$ is included analogously to the 2D case, leading to the following total Hamiltonian:
{\small \begin{equation}
    H^{(3\mathrm{D})} = \begin{pmatrix}
    \frac{E_g}{2} + H_{dd}^{(3\mathrm{D})}(\boldsymbol{k},\tfrac{\pi}{a_z}+k_z) & H_{dp}(\vec{\Delta}) \\
    H_{dp}^\dagger(\vec{\Delta}) & -\frac{E_g}{2} + H_{pp}^{(3\mathrm{D})}(\boldsymbol{k},k_z)
    \end{pmatrix},
\label{eq:Hkp_3D}
\end{equation}}
where $H_{dp}(\vec{\Delta})$ is defined by Eq.~\eqref{eq:HDelta}, and the $\tfrac{\pi}{a_z}$ shift of $k_z$ in the conduction band Hamiltonian ensures that the $L_1^-$ order parameter couples $\Gamma$ to $L$ (and $A$ to $M$). The in-plane dispersion for selected $k_z$ is provided in Fig. \ref{fig:3D_bands}(a), both in the normal state and in the CDW state with $|\vec{\Delta}|={82}\mathrm{meV}$, with a positive gap $E_g = +50 \mathrm{meV}$. The $k_z$ dispersion at the $\Gamma A$ and $ML$ directions is shown in Fig. \ref{fig:3D_bands}(b). Fig. \ref{fig:3D_bands}(c) displays the resulting Fermi surface when the Fermi energy $\mu=75\mathrm{meV}$ intersects the conduction band around the $L$ points. Fig. \ref{fig:3D_bands}(d) shows the DOS per unit cell for fixed $\theta_{d}=60\degree$ and varying $4 c_{d} = E_d(M) - E_d(L)$ from 60 to 210 meV. The monolayer limit is also shown for comparison, which would correspond to $\theta_{d}=90\degree$ and $c_{d}=0$. While there is no longer a peak and the its overall value is reduced, the DOS in 3D displays a significant jump at the same energy where the incipient saddle point develops in the 2D case, signaling the quasi-2D character of the bands. 

\subsection{Ground state in the 3D $\mathbf{k}\cdot\mathbf{p}$ model}

\begin{figure}[!t]
    \centering
    \includegraphics[width=\linewidth]{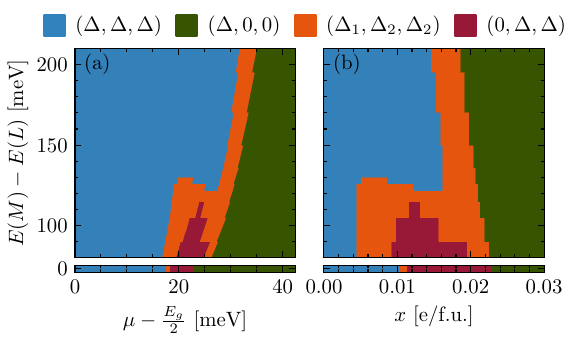}
    \caption{CDW ground state for the 3D $\boldsymbol{k}\cdot\boldsymbol{p}$ model of Eq.~\eqref{eq:Hkp_3D} for $\theta_d=60\degree$ and $|\vec{\Delta}|={82}\mathrm{meV}$ as a function of the chemical potential (a) or electron doping (b) and the energy difference between the electron pockets at $M$ and $L$, $E(M)-E(L) = 4c_d$. The bottom row indicates the ground state in the 2D case, where $E(M)-E(L)=0$ and $\theta_d=90\degree$.} 
    \label{fig:3D_phase_diagram}
\end{figure}

Analogously to the 2D case, we determine the CDW ground state for a fixed magnitude of $\vec{\Delta}$ for the 3D $\boldsymbol{k}\cdot\boldsymbol{p}$ model of Eqs. (\ref{eq:H3D_ini}-\ref{eq:H3D_fin}). Fig. \ref{fig:3D_phase_diagram}(a) shows the resulting phase diagram as a function of the chemical potential and $4 c_{d} = E_d(M) - E_d(L)$ for $\theta_{d}=60\degree$ and $E_g=50\mathrm{meV}$. In the 2D case ($c_d=0$ and $\theta_d=90\degree$), the transition to the nematic CDW occurs around $\mu = \tfrac{1}{2} E_g + 18\mathrm{meV}$, with the nematic CDW being mainly $2Q$ within the $\boldsymbol{k}\cdot\boldsymbol{p}$ model. Doping above $\mu = \tfrac{1}{2} E_g + 23\mathrm{meV}$ induces the $1Q$ stripe phase. Remarkably, both the nematic and the stripe phases survive in the 3D case. With increasing $c_{d}$ and thus 3D character, the nematic phase is suppressed, and the transition to the stripe phase occurs at slightly higher energies. Two regions of nematic phase can be differentiated: one around the energy of the incipient saddle point in 2D, $\mu \sim \tfrac{1}{2} E_g + 20\mathrm{meV}$, and another just at the border of the transition to the $1Q$ stripe phase, whose onset chemical potential increases with $c_{d}$. The first region, which is completely suppressed when $4c_{d}$ increases above 130 meV, is therefore driven by the same mechanism as in the 2D case. In the second region, the order parameter can be parametrized by $\vec{\Delta} \propto (\alpha,1-\alpha,1-\alpha)$, with $\alpha\in[0,1]$ such that $\alpha$ monotonically increases from 0.5 [$\vec{\Delta} \propto 0.5 (1,1,1)$] to 1 [$\vec{\Delta} \propto (1,0,0)$]. This indicates that, while it remains first order, the $C_3$-symmetric $3Q$ to stripe $1Q$ transition in 3D is less abrupt than in 2D, with an intermediate interpolating $3Q$ nematic phase. 

It is instructive to analyze the same phase diagram as a function of electron doping per unit cell, as shown by Fig. \ref{fig:3D_phase_diagram}(b). Interestingly, due to the different dependence between the chemical potential and the electron number in 2D and 3D, the nematic CDW onsets at a smaller electron doping and extends over a wider range of dopings in 3D for $4c_d \lesssim 130 \mathrm{meV}$. We finally mention that, for smaller $\theta_d$, or equivalently larger $d_d$, the nematic dome survives at higher $c_d$ (see Fig.~\ref{Sfig:3D_phase_diagram_c} in App.~\ref{app:3D_kp_model}). 

In conclusion, the doping-induced transitions to nematic and stripe phases within the CDW survive in bulk TiSe$_2$. In particular, the nematic CDW is also stable in 3D thanks to the combination of three facts: (1) the relatively strong coupling character of the CDW, illustrated by the large $|\vec{\Delta}|$ compared to the chemical potential where the nematic phase onsets; (2) the layered nature of TiSe$_2$, implying relatively small out-of-plane dispersion ($c_{p/d}$); and (3) the significant tilting of the conduction band Fermi surface with respect to the out-of-plane axis, leading to a relatively large $d_d$. Furthermore, due to the different dependence of doping with chemical potential in 3D, the doping where the nematic phase onsets is significantly reduced in 3D. This observation further supports the scenario where the growth-dependent native doping present in TiSe$_2$ samples \cite{Campbell19} might affect the symmetry of the CDW, inducing the nematic phase, as in Refs.~\cite{ILS10,Ishioka11,Kim24seamless}.

\section{Discussion} 
\label{sec:discussion}

In this work, we have argued that conduction band doping can lead to a 3Q nematic phase in TiSe$_2$ at low temperatures which is consistent with the STM observations in Refs.~\cite{ILS10,Ishioka11,Iavarone12,Kim24seamless}. Without the need to invoke an extra order parameter, our work explains the origin of $C_3$ anisotropy as a secondary instability of the CDW conduction bands in the presence of experimentally realistic ellipticity. {While formally this state breaks both translational and $C_3$ symmetries, we refer to it as nematic because it always originates from the $C_3$-symmetric 3Q state through a secondary nematic transition where only $C_3$ symmetry is broken.} A main point of this work is to emphasize that there is a natural explanation for the breaking of $C_3$ which does not require breaking inversion symmetry {with additional order parameters}, a key signature of the chiral CDW proposal ~\cite{vanWezel11,van2012chiral,Zenker13,GW15,Peng22,Subedi22}. 

Several experiments can be used to confirm the predictions of our work. First, if there is a nematic transition, it should be seen in the nematic susceptibility through a sharp change in elastoresistance. This might have already been detected below 200 K in the early days~\cite{Nunex83}, but we hope new experiments can be conducted to confirm it. $C_3$-breaking is also observable in low frequency Raman as the splitting of E modes, as observed in the orthorhombic phase of 2H-TaSe$_2$~\cite{Scott83}. {Notably, a recent Raman experiment has indeed seen this splitting \cite{kim2024origin}, which represents bulk evidence of $C_3$ breaking beyond STM. ARPES experiments may also detect the breaking of $C_3$ symmetry by comparing the ARPES intensity at the three $L$ points. In the nematic $3Q$ state, the three conduction band pockets are folded to each $L$ point and hybrdized, and only one is occupied and detectable in ARPES. However, the weight of this pocket will be different at each $L$ point, signaling the breaking of $C_3$ symmetry. Given the nematic state has three possible domains, such experiments should be performed with a small beam spot like in laser ARPES, so that a single domain can be resolved and the anisotropy features do not average out. Alternatively, ARPES on slightly strained samples may be performed to choose a single domain. While the $C_3$ symmetry is broken externally, the temperature dependence of the ARPES intensity at the three $L$ points should still show a crossover where one of the pockets becomes significantly brighter than the others.}

{While our candidate phase to explain the STM experiments is the nematic $3Q$ state, our work also predicts a $1Q$ stripe phase at high doping. Since the transition to the $1Q$ state is first order, there may be phase coexistence between the $1Q$ and $3Q$ phases. This may be a plausible explanation for the coexistence of $1Q$ and $3Q$ short-range domains observed in Cu-doped~\cite{Novello17} and Ta-doped \cite{Hu24Stripe} samples. However, it should be born in mind that such local domains could also be induced by the symmetry breaking due to the local dopants themselves, without the $1Q$ phase being a properly stable phase. More work is therefore needed to establish whether a global $1Q$ phase has really been observed. }

{Finally, as mentioned in Sec.~\ref{subsec:intro_sym_break}, nonlinear transport experiments like SHG and PGE represent ideal probes to determine if inversion is broken}. The only experiment showing such behavior used bulk samples cooled in the presence of circularly polarized light, which displayed a longitudinal circular photogalvanic effect consistent with preserved $C_3$ but broken inversion symmetry ~\cite{Xu20,jog_optically_2023}. {This suggests the possibility of a metastable state induced by the light training which is different from the one observed in STM, and it does display true chirality.} This state is likely induced by the condensation of subleading phonons with $M_1^-$ symmetry~\cite{Wickramaratne22} {(see also Table~\ref{tab:symmetry_groups_monolayer})}, or a mix of  $L_1^-$ and $M_1^-$ \cite{Subedi22,Kim24seamless}, aided by circularly polarized light. By considering the coupling to light in our model, our theory could be generalized to address this case.

In summary, the main conclusion of this work is that, taking into account all experimental evidence, it appears more likely that an inversion preserving $L_1^-$ nematic state is realized in TiSe$_2$. Our work provides microscopic arguments to explain the origin of such state, but we hope further experimental work will contribute to support its existence and settle the important question of the symmetry of the CDW in TiSe$_2$.

\section{Acknowledgements}
We are grateful to M. Ugeda, M. Gastiasoro, F. Flicker, J. van Wezel and M. Watson for interesting discussions on our work. D.M.S. acknowledges support from Spanish Ministerio de Ciencia, Innovaci\'on y Universidades (MCIU) FPU fellowship No. FPU19/03195{, and from the National Science Foundation (NSF) Materials Research Science and Engineering Centers (MRSEC) program through Columbia University under the Precision-Assembled Quantum Materials (PAQM) Grant No. DMR-2011738}. A.G.G. acknowledges financial support from the European Research Council (ERC) Consolidator grant under grant agreement No. 101042707 (TOPOMORPH). J.W.F.V. was supported by the National Science Foundation (NSF) Award No. DMR-2144352. F.J. acknowledges funding from the Spanish MCI/AEI/FEDER through grant PID2021-128760NB-I00.

\section*{Appendix}
\appendix

\section{Group theory, character tables and representation matrices}
\label{app:group_theory}

The normal state of TiSe$_2$ has the symmorphic space group $P \bar{3}m1$ ($\#164$), with point group $D_{3d}$. We will only consider the symmetry groups without spin. Its generators are $\{C_{3z}, C_{2x}, i\}$, with the center located in a Ti site. Its irreducible representations (irreps) are $\{ A_{1g} \equiv \Gamma_1^+, A_{2g} \equiv \Gamma_2^+, E_{g} \equiv \Gamma_3^+, A_{1u} \equiv \Gamma_1^-, A_{2u} \equiv \Gamma_2^-, E_{u} \equiv \Gamma_3^-\}$ (see \cite{aroyo_crystallography_2011} for the character table). The subindex $1/2$ in the $A$ irreps indicates the parity under $C_{2x}$, and the subindex $g/u$ or the superindex $\pm$ refers to the parity under the intralayer inversion $i$.

The commensurate CDW has wavector $L$ in the bulk and $M$ in the monolayer. Their little cogroup is $C_{2h}$. With the choice of the three $Q = \Gamma L, \Gamma M$ of Fig.~1 of the main text, $\boldsymbol{Q}_1 = (0,\frac{2\pi}{\sqrt{3}},Q_z)$, $\boldsymbol{Q}_2 = (-\frac{\sqrt{3}}{2} \frac{2\pi}{\sqrt{3}} , -\frac{1}{2} \frac{2\pi}{\sqrt{3}},Q_z)$, $\boldsymbol{Q}_3 = (\frac{\sqrt{3}}{2} \frac{2\pi}{\sqrt{3}} , -\frac{1}{2} \frac{2\pi}{\sqrt{3}},Q_z)$, where $Q_z = \pi, 0$, the little cogroup of $\boldsymbol{Q}_1$ has generators $\{C_{2x},i\}$. Their irreps are $\{Q_{1/2}^{\pm}\}$, which are all one-dimensional. Therefore, since there are three symmetry-equivalent $Q$ points in the star, the order parameters with wavevector $Q$ are three-dimensional.

A convenient approach to deal with the symmetry classification of $Q = \Gamma L, \Gamma M$ instabilities is the so-called extended point group \cite{venderbos_symmetry_2016,venderbos_multi-q_2016}, where the translations that are broken by the CDW are included in the point group. Effectively one determines the symmetry group of the $2\times2(\times2)$ supercell, and classifies the observables according to the irreps of this extended point group. 

\begin{table*}
	\begin{center}
    \caption{Character table of the extended point group $D_3^{(M)}$, and one-to-one correspondence with the cubic point group $O$.}	
    \label{table:character_table_extended_group}
	\begin{tabularx}{1\linewidth}{LLLLLL}
		\hline \hline
		& $1E$ 	& $3t (\equiv 3 C_{2[100]})$ 	& $6 C_2 (\equiv 6 C_{2[1-10]})$ 	& $8 C_3$ & $6 t C_2 (\equiv 6 C_4)$ 	\\ \hline
		$\Gamma_1 (\equiv A_1)$	& 1		& 1		& 1		& 1 	& 1		\\ 
		$\Gamma_2 (\equiv A_2)$	& 1		& 1		& -1	& 1 	& -1	\\ 
		$\Gamma_3 (\equiv E)$	& 2		& 2		& 0		& -1	& 0		\\ 
		$M_1 (\equiv T_2)$		& 3		& -1	& 1		& 0		& -1	\\ 
		$M_2	(\equiv T_1)$	& 3		& -1	& -1	& 0		& 1	\\ \hline \hline
	\end{tabularx}
	\end{center}

	\begin{center}
    \caption{Representation matrices of the generators of the extended point group $D_3^{(M)}$.}	
	\label{table:representation_matrices_extended_group}
	\begin{tabularx}{1\linewidth}{LLLLL}
		\hline \hline
		            & $t_1$     & $t_2$     & $C_{3z}$      & $C_{2x}$	\\ \hline
		$\Gamma_1$	& 1		& 1		& 1		& 1 			\\ 
		$\Gamma_2$	& 1		& 1		& 1	    & -1 		\\ 
		$\Gamma_3$	& $\begin{pmatrix} 1 & 0 \\ 0 & 1 \end{pmatrix}$
                    & $\begin{pmatrix} 1 & 0 \\ 0 & 1 \end{pmatrix}$
                    & $\begin{pmatrix} -\tfrac{1}{2} & -\tfrac{\sqrt{3}}{2} \\ \tfrac{\sqrt{3}}{2} & -\tfrac{1}{2} \end{pmatrix}$ 
                    & $\begin{pmatrix} 1 & 0 \\ 0 & -1 \end{pmatrix}$ \\ 
		$M_1$     & $\begin{pmatrix} 1 & 0 & 0 \\ 0 & -1 & 0 \\ 0 & 0 & -1 \end{pmatrix}$
                  & $\begin{pmatrix} -1 & 0 & 0 \\ 0 & 1 & 0 \\ 0 & 0 & -1 \end{pmatrix}$ 
	              & $\begin{pmatrix} 0 & 0 & 1 \\ 1 & 0 & 0 \\ 0 & 1 & 0 \end{pmatrix}$ 
	              & $\begin{pmatrix} 1 & 0 & 0 \\ 0 & 0 & 1 \\ 0 & 1 & 0 \end{pmatrix}$ \\ 
		$M_2$	  & $\begin{pmatrix} 1 & 0 & 0 \\ 0 & -1 & 0 \\ 0 & 0 & -1 \end{pmatrix}$ 
	              & $\begin{pmatrix} -1 & 0 & 0 \\ 0 & 1 & 0 \\ 0 & 0 & -1 \end{pmatrix}$ 
	              & $\begin{pmatrix} 0 & 0 & 1 \\ 1 & 0 & 0 \\ 0 & 1 & 0 \end{pmatrix}$ 
	              & $\begin{pmatrix} -1 & 0 & 0 \\ 0 & 0 & -1 \\ 0 & -1 & 0 \end{pmatrix}$ \\ \hline \hline
	\end{tabularx}
	\end{center}
\end{table*}

For the monolayer, where $Q = \Gamma M$, we perform group multiplication of the original point group $D_{3d}$ with the group $\{E, t_1, t_2, t_3\}$, where $t_1$ represents the translation by $\boldsymbol{a}_1 = a(1,0)$, $t_2$ by $\boldsymbol{a}_2-\boldsymbol{a}_1 = a(-\tfrac{1}{2},\tfrac{\sqrt{3}}{2})$, and $t_3$ by $-\boldsymbol{a}_2 = a(-\tfrac{1}{2},-\tfrac{\sqrt{3}}{2})$. Due to the imposed translational symmetry with a $2\times2$ unit cell, the group multiplication rules are $t_i t_i = E$ and $t_i t_j = t_k$, with $i \neq j \neq k$. The extended point group in the monolayer, $D_{3d}^{(M)} = D_{3d} \wedge \{ E, t_1, t_2, t_3\}$, is isomorphic to the cubic point group $O_h$. The generators of $D_{3d}^{(M)}$ are $\{C_{3z}, C_{2x}, i, t_1, t_2\}$, which are related to the generators $\{C_{3[111]}, C_{2[1-10]}, i, C_{2[001]}, C_{2[100]}\}$ of $O_h$ via the isomorphism. The character table of $D_{3}^{(M)}$ (from which $D_{3d}^{(M)}$ is obtained by the direct product with the intralayer inversion $i$) and its correspondence with the point group $O$ is shown in table~\ref{table:character_table_extended_group}. The classes are $3 t = \{t_1,t_2,t_3 \}$, $6 C_2 = \{ C_{2l}, t_l C_{2l} \}$ ($C_2$ rotations and products of translations along axis $l$ times $C_2$ rotations with the same axis $l$), $8 C_3 = \{ C_{3z}^{\pm}, t_l C_{3z}^{\pm} \}$, and $6 t C_2 = \{t_l C_{2m} \}$ (products of translations along axis $l$ times $C_2$ rotations with different axis $m$). {The intuition behind the character table of $D_{3}^{(M)}$ is the following. The space group irreps at the $\Gamma$ point are the 1D $\Gamma_1$ and $\Gamma_2$ and the 2D $\Gamma_3$, which correspond to the $A_1$, $A_2$ and $E$ irreps of the point group $D_{3}$, and classify the $\boldsymbol{q}=0$ orders. The little group at $M$ is $C_2$ with time reversal symmetry, which has two 1D real irreps, one even and one odd under $C_2$. The irreps of the little group at $M$ acquire a phase $\exp[i\boldsymbol{M}\cdot\boldsymbol{a}]$ under the translations by a original lattice vector $\boldsymbol{a}$. $C_3$ imposes the degeneracy between the irreps at the three $M$ points. Therefore, there are two 3D real space group irreps at the $M$ points, which classify the $\boldsymbol{q}=\boldsymbol{M}$ orders. When the intralayer inversion symmetry $i$ is included, these two irreps are further characterized by the parity under $i$.} The representation matrices of the generators of $D_{3}^{(M)}$ are shown in table~\ref{table:representation_matrices_extended_group}. As detailed in App.~\ref{app:kp_construction}, the character table and the representation matrices allow us to build the symmetry constrained $\boldsymbol{k} \cdot \boldsymbol{p}$ Hamiltonian and its coupling to the order parameter.

In the bulk, $Q= \Gamma L$, and the extended point group $D_{3d}^{(L)}$ is the direct product of $D_{3d}^{(M)}$ with the interlayer inversion symmetry $I$, with center in the midpoint between two Ti sites in adyacent layers, which commutes with all the other symmetry operations. Therefore, the irreps just carry an additional label indicating the parity under this interlayer inversion.

{As explained in Sec.~\ref{sec:phenomenology} of the main text, the special feature of $2\times2(\times2)$ CDWs is that the irreps, and therefore the order parameters, characterizing them are real, and not complex. This is reproduced by our symmetry analysis, where the $M_i^\pm$ irreps are real. Another way to see this is that for general commensurate states, the order parameter at $\boldsymbol{Q}_i$ transforms under a lattice translation $\boldsymbol{a}$ as $e^{i \boldsymbol{Q}_i \cdot \boldsymbol{a}}$ which necessarily requires the order parameter to be complex. But when $\boldsymbol{Q}_i = \tfrac{1}{2}  \boldsymbol{G}_i$ these translation operators become real, and the order parameter should be taken real.}

\section{k.p model}

In this section, we first explain the procedure to derive the $\boldsymbol{k}\cdot\boldsymbol{p}$ model of Eqs.~(\ref{eq:H_dd}-\ref{eq:Hpp0}) of the main text by using the symmetries of monolayer TiSe$_2$ provided in Appendix \ref{app:group_theory}. Then, we analytically determine the critical doping for the $3Q$ to $1Q$ transition in the isotropic limit where $b_p=b_d=0$. Finally, we extend the $\boldsymbol{k}\cdot\boldsymbol{p}$ model to 3D.

\subsection{Construction of the k.p by symmetry} \label{app:kp_construction}

The low-energy band structure of monolayer TiSe$_2$ is sketched in Fig.~\ref{Fig1} of the main text. The two band eigenstates at $\Gamma$ transform as $\Gamma_3^-$, which corresponds to the $E_u$ irreducible representation (irrep) of the little group $D_{3d}$. We define their annihilation operators $p = \{p_x,p_y\}$. We can now classify their Hermitian fermion bilinears $p\dag p$ according to symmetry as:
\begin{align}
    \Gamma_1^+\left(A_{1g}\right) \rightarrow & p_x^\dagger p_x + p_y^\dagger p_y, \\
    \Gamma_2^+\left(A_{2g}\right) \rightarrow & -i\left(p_x^\dagger p_y - p_y^\dagger p_x\right), \label{Seq:pp_A2g}\\
    \Gamma_3^+\left(E_g\right) \rightarrow & \left\{p_x^\dagger p_x - p_y^\dagger p_y, -\left(p_x^\dagger p_y + p_y^\dagger p_x \right) \right\},
\end{align}
where the notation in parenthesis corresponds to the little group irreps. Now, crystal momentum $\boldsymbol{k} = (k_x,k_y)$ transforms as $E_u$ in $D_{3d}$, and therefore its quadratic combinations transform as:
\begin{align}
    A_{1g} \rightarrow & k_x^2+k_y^2 \coloneqq k^2, \label{Seq:k_A1g}\\
    E_g \rightarrow & \left\{k_x^2-k_y^2, -2k_xk_y \right\}, \label{Seq:k_Eg}
\end{align}
where the $A_{2g}$ combination $k_x k_y - k_y k_x$ vanishes since it is antisymmetric. The symmetry-allowed $A_{1g}$ combinations of $p$ bilinears and momenta enter the $H_{pp}^0$ Hamiltonian of Eq.~\eqref{eq:Hpp0} of the main text.

Now, we derive the Hamiltonian $H_{dd}^0$ of the three electron pockets at the $M$ points, which transform as $M_1^+$ ($A_{1g}$ of the little group $C_{2h}$). We define their annihilation operators $d = \{d_1,d_2,d_3\}$, where $d_i$ represents the band at $\boldsymbol{M}_i$, with $\boldsymbol{M}_1 = \tfrac{2\pi}{\sqrt{3}}(0,1)$, $\boldsymbol{M}_2 =  \tfrac{2\pi}{\sqrt{3}}(-\tfrac{\sqrt{3}}{2}  , -\tfrac{1}{2} )$, $\boldsymbol{M}_3 =  \tfrac{2\pi}{\sqrt{3}}(\tfrac{\sqrt{3}}{2}  , -\tfrac{1}{2} )$ (see Fig.~\ref{Fig1} of the main text). In the normal state, the three electron bands are uncoupled. Therefore, we could derive the energy of, for example, $d_1$, and then obtain the dispersions of $d_2$ and $d_3$ by applying the threefold rotational symmetry connecting the three $M$ points. Alternatively, we derive $H_{dd}^0$ in the $2\times2$ CDW supercell using the extended point group introduced in Appendix \ref{app:group_theory}. The $d\dag d$ bilinears transform under the symmetry of the extended point group as
\begin{align}
    \Gamma_1^+ \left(A_{1g}\right) \rightarrow & d\dag_1 d_1 + d\dag_2 d_2 + d\dag_3 d_3, \label{Seq:A1g_d_bands}\\
    \Gamma_3^+ \left(E_{g}\right) \rightarrow & \left\{ d\dag_1 d_1 - \tfrac{1}{2} d\dag_2 d_2 - \tfrac{1}{2} d\dag_3 d_3, \tfrac{\sqrt{3}}{2} \left( d\dag_2 d_2 - d\dag_3 d_3 \right) \right\}, \label{Seq:Eg_d_bands} \\
    M_1^+ \rightarrow & \left\{ d\dag_2 d_3 + d\dag_3 d_2, d\dag_3 d_1 + d\dag_3 d_1, d\dag_1 d_2 + d\dag_2 d_1 \right\}, \\
    M_2^+ \rightarrow & i \left\{ d\dag_2 d_3 - d\dag_3 d_2, d\dag_3 d_1 - d\dag_3 d_1, d\dag_1 d_2 - d\dag_2 d_1 \right\}.
\end{align}
The combinations of \eqref{Seq:A1g_d_bands} and \eqref{Seq:Eg_d_bands} with the crystal momentum terms \eqref{Seq:k_A1g} and \eqref{Seq:k_Eg}, respectively, lead to the Hamiltonian $H_{dd}^0$ of Eqs.~(\ref{eq:H_dd}-\ref{eq:Ed3}).

Finally, we consider the coupling to the CDW order parameter with $M_1^-$ symmetry and even under time-reversal symmetry, denoted $\vec{\Delta} = (\Delta_1,\Delta_2,\Delta_3)$, which hybridizes the valence and conduction bands. In order to derive the form of the coupling, we need the symmetry classification of the $d\dag p$ bilinears in the extended point group:
\begin{align}
    M_1^- \rightarrow & \left\{ d\dag_1 p_x, d\dag_2 \left( -\tfrac{1}{2} p_x + \tfrac{\sqrt{3}}{2} p_y \right), d\dag_3 \left( -\tfrac{1}{2} p_x - \tfrac{\sqrt{3}}{2} p_y \right) \right\}, \label{Seq:dp_M1}\\
    M_2^- \rightarrow & \left\{ d\dag_1 p_y, d\dag_2 \left( -\tfrac{1}{2} p_y - \tfrac{\sqrt{3}}{2} p_x \right), d\dag_3 \left( -\tfrac{1}{2} p_y + \tfrac{\sqrt{3}}{2} p_x \right) \right\}. \label{Seq:dp_M2}
\end{align}
Each $M_i^-$ bilinear allows for two Hermitian combinations: the time-reversal (TRS) even $(d\dag p)_{M_i^-} + \mathrm{h.c.}$, and the time-reversal odd $i[(d\dag p)_{M_i^-} - \mathrm{h.c.}]$. The time-reversal-symmetric $M_1^-$ order parameter $\vec{\Delta}$ couples to the time-reversal even combination of \eqref{Seq:dp_M1}. To lowest order in $|\vec{\Delta}| = \sqrt{\Delta_1^2+\Delta_2^2+\Delta_3^2}$ and $\boldsymbol{k}$, the symmetry-allowed coupling to $\vec{\Delta}$ is therefore described by the Hamiltonian of Eqs.~(\ref{eq:H_kp}-\ref{eq:HDelta}) of the main text. 

As deduced from Eqs.~(\ref{Seq:dp_M1}, \ref{Seq:dp_M2}), three other order parameters would also couple the electron and hole pockets: a TRS odd $M_1^-$, a TRS even $M_2^-$, and a TRS odd $M_2^-$. In the isotropic $b_p=b_d=0$ case, the four order parameters have exactly the same effect on the band eigenvalues. Away from this limit, the $M_1^-$ and $M_2^-$ order parameters differ, but the time reversal even and odd combinations are still degenerate. As we will discuss in Appendix \ref{app:TB}, this degeneracy arises due to the artificial $U(1)$ symmetry of the $\boldsymbol{k}\cdot \boldsymbol{p}$ model representing the separate charge conservation in the valence and conduction bands in the normal state. In the $\boldsymbol{k}\cdot \boldsymbol{p}$ model, this degeneracy can only be broken by adding umklapp interactions \cite{ganesh_theoretical_2014}, but it is naturally lifted in a tight-binding model. Here we consider only the coupling to the time-reversal even $M_1^-$ order parameter in the $\boldsymbol{k}\cdot \boldsymbol{p}$ model, which is the only one of the four that condenses in TiSe$_2$.

\subsection{Analytical solution for the 3\textit{Q}-1\textit{Q} critical doping}
\label{app:kp_energy_analytic}

The existence of a critical doping $x_{1Q}$ above which a $1Q$ solution for $\vec{\Delta}$ is obtained can be shown analytically in the $\boldsymbol{k} \cdot \boldsymbol{p}$ model in the simplified case where $b_p = b_d=0$. 

Consider a generic state $\vec{\Delta} = (\Delta_1,\Delta_2,\Delta_3)$ parametrized as $\vec{\Delta} = |\vec{\Delta}| [\cos(\theta), \sin(\theta) \cos(\varphi), \sin(\theta) \sin(\varphi)]$ with $|\vec{\Delta}| = \sqrt{\Delta_1^2+\Delta_2^2+\Delta_3^2}$, $\theta\in[0,\pi]$, and $\varphi\in[0,2\pi]$. The energies of the bands for $b_p = b_d=0$ are:
{\small
\begin{align}
    \varepsilon_{c0}(\boldsymbol{k}) &= \frac{E_g}{2}+a_d k^2, \\
    \varepsilon_{c \pm}(\boldsymbol{k}) &= \frac{1}{2}\left[a_{dp+}k^2 + \sqrt{\left(E_g+a_{dp-}k^2\right)^2 + {2} |\vec{\Delta}|^2 f_{\pm}(\theta,\varphi)} \right], \\
    \varepsilon_{v \pm}(\boldsymbol{k}) &= \frac{1}{2}\left[a_{dp+}k^2 - \sqrt{\left(E_g+a_{dp-}k^2\right)^2 + {2} |\vec{\Delta}|^2 f_{\pm}(\theta,\varphi)} \right],
\end{align}}
where $k = |\boldsymbol{k}|$ and we have defined:
\begin{align}
    & a_{dp\pm}=a_d\pm a_p, \\        
    & |\vec{\Delta}|^2 f_{\pm}(\theta,\varphi) = \left(|\vec{\Delta}|^2 \pm \sqrt{\tfrac{3}{2}(\Delta_1^4+\Delta_2^4+\Delta_3^4) - \tfrac{1}{2} |\vec{\Delta}|^4} \right).
\end{align}
$\varepsilon_{v \pm}$ are the two valence bands, $\varepsilon_{c 0}$ is the lowest conduction band, which always remains uncoupled, and $\varepsilon_{c \pm}$ are the two highest conduction bands.

Note that the energies only depend on the direction of $\vec{\Delta}$ via the quartic invariant $(\Delta_1^4+\Delta_2^4+\Delta_3^4)$, which, for a given modulus $|\vec{\Delta}|$, is minimum for $\vec{\Delta} = \frac{|\vec{\Delta}|}{\sqrt{3}} (1,1,1)$, and maximum for $\vec{\Delta} = |\vec{\Delta}| (1,0,0)$. Let us assume that the CDW phase displays a gap between valence and conduction bands so that the valence bands are fully filled at stoichiometry $x=0$ (for which $E_g>0$ is a sufficient condition). Then the ground state in the undoped case $x=0$ has to be either the $3Q$ $C_3$-symmetric state $\vec{\Delta} = \frac{|\vec{\Delta}|}{\sqrt{3}} (1,1,1)$ or the $1Q$ stripe state $\vec{\Delta} = |\vec{\Delta}| (1,0,0)$. For the $3Q$ $C_3$-symmetric state, two pairs of valence and conduction bands are repelled and remain degenerate:
{\small
\begin{align}
    \varepsilon_{c \pm}^{3Q}(\boldsymbol{k}) &=\varepsilon_{c}^{3Q}(\boldsymbol{k}) = \frac{1}{2}\left[a_{dp+}k^2 + \sqrt{\left(E_g+a_{dp-}k^2\right)^2+{2} |\vec{\Delta}|^2} \right], \\
    \varepsilon_{v \pm}^{3Q}(\boldsymbol{k}) &= \varepsilon_{v}^{3Q}(\boldsymbol{k}) = \frac{1}{2}\left[a_{dp+}k^2 - \sqrt{\left(E_g+a_{dp-}k^2\right)^2+{2} |\vec{\Delta}|^2} \right],
\end{align}}
while for the $1Q$ stripe state, only one pair of valence and conduction bands is repelled:
\begin{align}
    \varepsilon_{c +}^{1Q}(\boldsymbol{k}) &= \frac{1}{2}\left[a_{dp+}k^2 + \sqrt{\left(E_g+a_{dp-}k^2\right)^2 + {4} |\vec{\Delta}|^2} \right], \\
    \varepsilon_{c -}^{1Q}(\boldsymbol{k}) &= \varepsilon_{c0}(\boldsymbol{k}), \\
    \varepsilon_{v +}^{1Q}(\boldsymbol{k}) &= \frac{1}{2}\left[a_{dp+}k^2 - \sqrt{\left(E_g+a_{dp-}k^2\right)^2 + {4} |\vec{\Delta}|^2} \right], \\
    \varepsilon_{v-}^{1Q}(\boldsymbol{k}) &= -\frac{E_g}{2} + a_p k^2, 
\end{align}

In order to determine the ground state, let us compute the difference in total energy density at zero temperature $\delta E = E-E^{3Q}$ between a generic state $\vec{\Delta} = (\Delta_1,\Delta_2,\Delta_3)$ and the $\vec{\Delta} = \frac{|\vec{\Delta}|}{\sqrt{3}} (1,1,1)$ phase. At stoichiometry $x=0$,
\begin{widetext}
\begin{equation}
\begin{split}
    & \delta E (x=0) = \int \frac{d^2 k}{(2\pi)^2} \left[\varepsilon_{v+}(\boldsymbol{k}) + \varepsilon_{v-}(\boldsymbol{k}) - 2\varepsilon_v^{3Q}(\boldsymbol{k})\right] = \\
    &= \frac{1}{8 \pi (a_d-a_p)} \left\{ \frac{E_g}{2} \left[ \sqrt{E_g^2+{2}|\vec{\Delta}|^2 f_+(\theta,\varphi)} 
    + \sqrt{E_g^2+{2}|\vec{\Delta}|^2 f_-(\theta,\varphi)} 
    - 2\sqrt{E_g^2+{2}|\vec{\Delta}|^2} \right] + \right. \\
    & \left. + {2}|\vec{\Delta}|^2  \left[ \log{\left(-E_g + \sqrt{E_g^2+{2}|\vec{\Delta}|^2}\right)} - \frac{1}{2} \sum_{\alpha=\pm} f_\alpha(\theta,\varphi) \log{\left( \frac{-E_g + \sqrt{E_g^2+{2}|\vec{\Delta}|^2 f_\alpha(\theta,\varphi)}}{f_\alpha(\theta,\varphi)} \right)} \right] \right\}.
\end{split}
\end{equation}
$\delta E (x=0)$ is always a non-negative quantity, and it is equal to zero only if $\vec{\Delta} = \frac{|\vec{\Delta}|}{\sqrt{3}} (1,1,1)$, when $f_{\pm}(\theta,\varphi) = 1$. Therefore, the ground state at charge neutrality is the $3Q$ $C_3$-symmetric state, and the $1Q$ stripe state has the highest energy:
\begin{equation}
    \delta E^{1Q} (x=0) = \frac{1}{8 \pi (a_d-a_p)}\left[\frac{E_g}{2} \left(\sqrt{E_g^2+{4}|\vec{\Delta}|^2} + E_g
    - 2\sqrt{E_g^2+{2}|\vec{\Delta}|^2} \right) 
    + {2}|\vec{\Delta}|^2 \log{\left(2 \frac{-E_g + \sqrt{E_g^2+{2}|\vec{\Delta}|^2}}{-E_g + \sqrt{E_g^2+{4}|\vec{\Delta}|^2}}\right)} \right].
\end{equation}
\end{widetext}
In the limit $E_g \rightarrow 0$ we have simply $\delta E (x=0) = \tfrac{|\vec{\Delta}|^2}{{16}\pi(a_d-a_p)} [f_+ \log{(f_+)} + f_- \log{(f_-)}]$, and $\delta E^{1Q} (x=0) = \tfrac{|\vec{\Delta}|^2}{{8}\pi(a_d-a_p)}\log 2$. Let us note that $\delta E (x=0)$ decreases with increasing $E_g$ for fixed $|\vec{\Delta}|$.

Now consider doping a small carrier density $n = x/V_{\text{unit cell}}$ such that only the lowest conduction band $\varepsilon_{c0}$ is populated. This assumption holds in the majority of the $\boldsymbol{k}\cdot\boldsymbol{p}$ phase diagrams of Figs.~\ref{Fig2}(e,f) (except for some regions in the case of large negative gap and small ellipticity, as explained in the main text). This assumption is also verified in all the self-consistent mean-field calculations. Except for the $1Q$ phase, the lowest conduction band is non-degenerate and equal for all $|\vec{\Delta}|$, so that the total energy difference $\delta E (x)$ remains the same as $\delta E (x=0)$.

However, the $1Q$ stripe phase displays a doubly-degenerate lowest conduction band. In this case, for a given carrier density $n$, the chemical potential is lower for the $1Q$ state than for the $3Q$ one, which allows the possibility of a transition to the $1Q$ phase at a critical doping, as we show below. The lowest conduction band is uncoupled by the order parameter, and thus remains parabolic with constant DOS $1/(4\pi a_d)$. The chemical potential is set by the carrier density:
\begin{align}
    n = \int_{0}^{\mu^{3Q}} d\varepsilon \frac{1}{4\pi a_d} \theta(\varepsilon-E_g/2) = \frac{\mu^{3Q}-E_g/2}{4\pi a_d}\\   
    n = \int _{0}^{\mu^{1Q}} d\varepsilon \frac{2}{4\pi a_d} \theta(\varepsilon-E_g/2)= \frac{\mu^{1Q}-E_g/2}{2\pi a_d}\\  
\end{align}
And the total energy density difference is 
\begin{align}
\begin{split}
    & \delta E^{1Q} (x) - \delta E^{1Q} (x=0) = \\ 
    &= \int_{0}^{\mu^{1Q}} d\varepsilon \frac{\varepsilon}{2\pi a_d} \theta(\varepsilon-\tfrac{E_g}{2}) - \int_{0}^{\mu^{3Q}} d\varepsilon \frac{\varepsilon}{4\pi a_d} \theta(\varepsilon-\tfrac{E_g}{2}) = \\
    &= \frac{1}{8\pi a_d} \left[2(2\pi a_d n)^2 - (4\pi a_d n)^2 \right]  = - \pi a_d n^2 
\end{split}
\end{align}
The transition to the $1Q$ state occurs at the $x_{1Q}$ such that $\delta E^{1Q}(x_{1Q}) = 0$, so
\begin{align}
    n_{1Q} = 2 \sqrt{\frac{\delta E^{1Q}(x=0)}{\pi a_d}},
\label{Seq:nc_111_to_100}
\end{align}
where we have added a factor $2$ to take into account the spin degeneracy. In the limit $E_g \rightarrow 0$, we have $n_{1Q} = \tfrac{|\vec{\Delta}|}{\pi} \sqrt{\tfrac{\log 2}{{2}a_d(a_d-a_p)}}$.

We can estimate $x_{1Q}$ from this calculation taking $a_p = -0.89 \hbar^2/m_e$ and $a_d = 0.27 \hbar^2/m_e$, with $m_e$ the electron mass, which reproduce the same normal-state DOS as the realistic values $a_p = -0.95 \hbar^2/m_e$, $b_p = -0.24 \hbar^2/m_e$, $a_d = 0.54 \hbar^2/m_e$ and $b_d = 0.46 \hbar^2/m_e$ used in our effective tight-binding model. From the ARPES experiment on monolayer TiSe$_2$ of Ref.~\cite{Watson20}, where the normal-state gap is $E_g \sim 80 \mathrm{meV}$ and the low-temperature gap is $E_g/2+{\sqrt{(E_g/2)^2+|\vec{\Delta}|^2/{2}}} \sim 180 \mathrm{meV}$, one obtains an order parameter $|\vec{\Delta}| \sim {190} \mathrm{meV}$ in the low-doping case. Using these numerical values, we can estimate the critical doping for the transition from the $3Q$ to the $1Q$ states to be $x_{1Q} \sim 0.07 \mathrm{e/f.u.}$. Despite neglecting $b_p$ and $b_d$, this value is of the order of magnitude of that obtained in the self-consistent mean-field calculations. The quantitative agreement is even better for a smaller $|\vec{\Delta}|$, which accounts for its decrease with increasing doping. The numerical results for $x_{1Q}$ for $E_g>0$ and nonzero $b_p$ and $b_d$ are displayed in Fig.~\ref{Fig2}(e), which demonstrates that if the CDW survives at high enough doping, a $1Q$ phase universally appears for any ellipticity and gap.

\section{k.p model and ground state in 3D} \label{app:3D_kp_model}

In this Appendix, we provide further details of the derivation of the 3D $\boldsymbol{k}\cdot\boldsymbol{p}$ model of Eqs.~(\ref{eq:H3D_ini}-\ref{eq:H3D_fin}). The first step is to introduce the $k_z$ dispersion to order $\mathcal{O}(|\boldsymbol{k}|^2)$. The momentum $k_z$ transforms as $A_{2u}$ in the point group $D_{3d}$, and the quadratic combinations as $k_z^2 \rightarrow A_{1g}$ and $\{-k_z k_y, k_z k_x\} \rightarrow E_g$. Using the symmetry analysis of App.~\ref{app:kp_construction}, the noninteracting 3D $\boldsymbol{k}\cdot\boldsymbol{p}$ Hamiltonian reads:
\begin{align}
    \varepsilon_{d1}^{k_z}(\boldsymbol{k},k_z) =& \varepsilon_{d1}(\boldsymbol{k}) + c_d k_z^2 - d_d k_z k_y, \\
    \varepsilon_{d2}^{k_z}(\boldsymbol{k},k_z) =& \varepsilon_{d2}(\boldsymbol{k}) + c_d k_z^2 + d_d k_z \left(\frac{1}{2} k_y + \frac{\sqrt{3}}{2} k_x \right), \\
    \varepsilon_{d3}^{k_z}(\boldsymbol{k},k_z) =& \varepsilon_{d3}(\boldsymbol{k}) + c_d k_z^2 + d_d k_z \left(\frac{1}{2} k_y - \frac{\sqrt{3}}{2} k_x \right), \\
    H_{dd}^{0 k_z}(\boldsymbol{k},k_z) =& {\rm diag}\left[\varepsilon_{d,1}^{k_z}(\boldsymbol{k},k_z),\varepsilon_{d,2}^{k_z}(\boldsymbol{k,k_z}),\varepsilon_{d,3}^{k_z}(\boldsymbol{k},k_z)\right], \\
    \begin{split}
        H_{pp}^{0 k_z}(\boldsymbol{k},k_z) =& H_{pp}^0(\boldsymbol{k}) + \\
        &+ \left(\begin{array}{cc}
    - c_p k_z^2 - d_p k_z k_y & -d_p k_z k_x \\
    -d_p k_z k_x & - c_p k_z^2 + d_p k_z k_y
    \end{array}
    \right),
    \end{split}
\end{align}
where $k_z$ is measured from the $\Gamma$ point for the valence bands and from the $L$ point for the conduction bands, $\varepsilon_{di}(\boldsymbol{k},k_z)$ were defined in Eqs.~(\ref{eq:Ed1}-\ref{eq:Ed3}) of the main text and $H_{pp}^0(\boldsymbol{k},k_z)$ is given by Eq.~\eqref{eq:Hpp0} of the main text. As in the main text, the $c_{p/d}$ terms represent the inverse valence/conduction band masses in the $z$ direction, while the $d_{p/d}$ terms couple the in-plane $k_{x,y}$ and out-of-plane $k_z$ dispersions, and therefore tilt the ellipsoidal Fermi surfaces with respect to the $z$ direction. Similar expressions would describe the valence bands at $A$ and the conduction bands at $M$.

\begin{figure}[!t]
    \centering
    \includegraphics[width=0.8\linewidth]{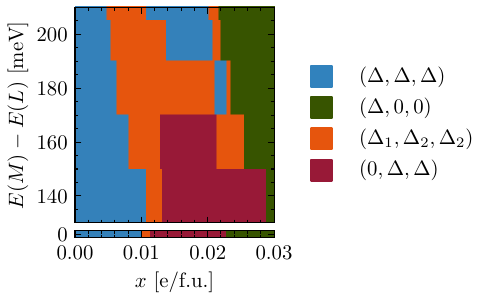}
    \caption{CDW ground state for the 3D $\boldsymbol{k}\cdot\boldsymbol{p}$ model of Eqs.~(\ref{eq:H3D_ini}-\ref{eq:H3D_fin}) for $\theta_d=45\degree$ and $|\vec{\Delta}| = {82} \mathrm{meV}$ as a function of the electron doping and the energy difference between the electron pockets at $M$ and $L$, $E(M)-E(L) = 4c_d$. The bottom row indicates the ground state in the 2D case, where $E(M)-E(L)=0$ and $\theta_d=90\degree$.}
    \label{Sfig:3D_phase_diagram_c}
\end{figure}

As pointed out in Sec.~\ref{sec:3D_kp_model}, the full out-of-plane dispersion in the Brillouin zone is well described by the lowest-order trigonometric functions respecting the symmetries. This is useful both for fitting the $c_{p/d}$ and $d_{p/d}$ parameters, as well as for comparing the DOS per unit cell between the 2D and 3D cases. On the one hand, the $c_{p/d} k_z^2$ terms can then be replaced by $2 c_{p/d} [1-\cos(a_z k_z)] = 4 c_{p/d} \sin^2(\tfrac{1}{2} a_z k_z)$, where $4 c_{p} = E_p(\Gamma) - E_p(A)$ corresponds to the energy difference between the valence bands at $\Gamma$ and $A$, and $4 c_{d} = E_d(M) - E_d(L)$ corresponds to the energy difference between the conduction bands at $M$ and $L$. On the other hand, the $d_{p/d} k_z$ terms can be replaced by $d_{p/d} \sin(a_z k_z)$, since these terms have to vanish both at $k_z=0$ and at $k_z=\pi$. The resulting $\boldsymbol{k}\cdot\boldsymbol{p}$ Hamiltonian capturing the full $k_z$ dispersion is given by Eqs.~(\ref{eq:H3D_ini}-\ref{eq:H3D_fin}) of the main text. 

For completeness, we present in Fig. \ref{Sfig:3D_phase_diagram_c} the phase diagram of the 3D $\boldsymbol{k}\cdot\boldsymbol{p}$ model of Eqs.~(\ref{eq:H3D_ini}-\ref{eq:H3D_fin}) for $\theta_d=45\degree$. It shows that the nematic dome survives at higher $c_d$ for smaller $\theta_d$, or equivalently larger $d_d$. We note that the y-axis of Fig. \ref{Sfig:3D_phase_diagram_c} is restricted to $4 c_{d} = E_d(M) - E_d(L) > 130 \mathrm{meV}$ since, below that threshold, the minimum of the conduction band no longer occurs at a high symmetry point.

\section{Tight-binding model and mean-field theory}
\label{app:TB}

Here we describe the effective tight-binding model, the choice of its parameters, and the details of the self-consistent mean-field calculations, {and we provide further phase diagrams obtained for different parameters}.

\subsection{Tight binding and model parameters} \label{app:TB_params}

The realistic tight-binding model of TiSe$_2$ would consist of at least 7 orbitals per unit cell: the four $\{p_x,p_y\}$ orbitals from the two Se atoms, which transform as $E_u$ representations, and the $t_{2g}$ triplet of $d$ orbitals $\{d_{xy},d_{yz},d_{zx}\}$ from the Ti (the approximately cubic environment of Ti makes it useful to refer to these orbitals as $t_{2g}$ even though they are actually split in a singlet $A_{1g}$ and a doublet $E_g$ because the overall crystal symmetry is trigonal).  In 2D and in the absence of SOC, the low-energy physics is dominated by two degenerate hole pockets at $\Gamma$ coming from the Se-$p$ orbitals which transform as the $\Gamma_3^-$ representation, and three electron pockets at the $M$ points coming from the Ti-$d$ orbitals which transform according to the $M_{1}^+$ representation. 

Here, we have considered instead an effective tight-binding model with the same space group $P \bar{3}m1$ (SG 164, PG $D_{3d}$) which reproduces the band dispersion and eigenstate symmetry near the Fermi level with 3 orbitals per unit cell (formally, the model also has a $m_z$ symmetry, but it plays no role because all orbitals are located at $z=0$). The 3 orbitals are located at the center of a triangular lattice. Two orbitals transform as $E_u$, and mainly compose the valence bands, so they will be denoted as $\{p_x,p_y\}$. The other orbital is totally symmetric ($A_{1g}$), and mainly composes the conduction band, so it will be denoted as $d_{z^2}$. The non-interacting model has 8 parameters: onsite energies $\varepsilon_p$ and $\varepsilon_d$, hoppings up to third nearest neighbours $t_{dd}^{(n)}$ for the $d_{z^2}$ orbital, $\sigma$ and $\pi$ nearest-neighbour hoppings $t_{pp\sigma}$ and $t_{pp\pi}$ for the $p$ orbitals, and nearest-neighbour hopping $t_{dp}$ coupling the $d$ and $p$ orbitals. The non-interacting Hamiltonian is given by Eq.~\eqref{eq:H0_TB} of the main text. In this Hamiltonian we have neglected SOC. While this can quantitatively change the critical temperature and related quantities, we expect that the qualitative picture remains the same~\cite{hellgren_critical_2017}.

We choose the Hamiltonian parameters by solving for the non-interacting gap $E_g$, the masses $m_{v1}, m_{v2}$ of the two valence bands at $\Gamma$, the masses $m_{cx}, m_{cy}$ of the conduction bands at $M$ perpendicular and parallel to the $\Gamma M$ direction, and the energy $\varepsilon_{c\Gamma}$ of the conduction band at $\Gamma$. The resulting system of equations would be underconstrained, so we choose to leave $t_{dp}$ as a free parameter. Since the valence bands are of $p$ character while the conduction bands are of $d$ character, $t_{dp}$ only affects the band curvatures to second order with an energy denominator dominated by $\varepsilon_d - \varepsilon_p$, so its influence on the bands is almost negligible. Nevertheless, we choose to keep it finite because it breaks the artificial $U(1)$ gauge symmetry representing the separate charge conservation in the conduction and valence bands. The subtle role of $t_{dp}$ in selecting a mean field solution is further explained below in Sec. \ref{app:MF_order_parameters}.

The gap, the masses and $\varepsilon_{c\Gamma}$ depend on the hoppings as:
\begin{widetext}
\begin{align}
	& E_g = \varepsilon_d - \varepsilon_p - \left[2 t_{dd}^{(1)} + 2 t_{dd}^{(2)} - 6 t_{dd}^{(3)} - 3 t_{pp\sigma} +3 t_{pp\pi} \right] \nonumber \\
	& \varepsilon_{c\Gamma} = \varepsilon_d + 6 \left[t_{dd}^{(1)} + t_{dd}^{(2)} + t_{dd}^{(3)} \right] \nonumber \\
    & \widetilde{m}_{v1}^{-1} = \frac{3}{4} \left[ t_{pp\sigma} - 3 t_{pp\pi} \right] \nonumber \\
    & \widetilde{m}_{v2}^{-1} = -\frac{3}{4} \left[ - 3 t_{pp\sigma} + t_{pp\pi} + \frac{24 t_{dp}^2}{\varepsilon_d - \varepsilon_p + 3 \left(2 t_{dd}^{(1)} + 2 t_{dd}^{(2)} + 2 t_{dd}^{(3)} + t_{pp\sigma} - t_{pp\pi} \right)} \right] \label{Seq:to_fit_simple} \\
    & \widetilde{m}_{cy}^{-1} = 3 \left[ t_{dd}^{(1)} - t_{dd}^{(2)} - 4 t_{dd}^{(3)} + \frac{6 t_{dp}^2}{\varepsilon_d - \varepsilon_p - \left( 2 t_{dd}^{(1)} + 2 t_{dd}^{(2)} - 6 t_{dd}^{(3)} + 3 t_{pp\sigma} + t_{pp\pi} \right)} \right] \nonumber \\
	& \widetilde{m}_{cx}^{-1} = - t_{dd}^{(1)} + 9 t_{dd}^{(2)} - 12 t_{dd}^{(3)} + \frac{2 t_{dp}^2}{\varepsilon_d - \varepsilon_p - \left(2 t_{dd}^{(1)} + 2 t_{dd}^{(2)} - 6 t_{dd}^{(3)} -  t_{pp\sigma} - 3 t_{pp\pi}\right)} \nonumber
\end{align}
where we have defined $\widetilde{m} = m a^2 / \hbar^2$, where $a$ is the lattice constant. Inverting these relationships, we find that the hoppings as a function of the gap, the masses, $\varepsilon_{c\Gamma}$ and $t_{dp}$ can be expressed as:
\begin{align}
	& \varepsilon_d = \frac{1}{256} \left(48 \widetilde{m}_{cx}^{-1}+80 \widetilde{m}_{cy}^{-1}+\frac{288 t_{dp}^2}{8 \widetilde{m}_{v1}^{-1}-3 E_g}-\frac{15 (8 \widetilde{m}_{v2}^{-1}-3 E_g) (2 \varepsilon_{c\Gamma}+E_g)^2}{(8 \widetilde{m}_{v2}^{-1}-3
   E_g) (2 \varepsilon_{c\Gamma}+E_g)+288 t_{dp}^2}+70 \varepsilon_{c\Gamma}+123 E_g\right) \nonumber \\
    & \varepsilon_p = \widetilde{m}_{v1}^{-1}+\widetilde{m}_{v2}^{-1}+\frac{36 t_{dp}^2}{2 \varepsilon_{c\Gamma}+E_g}-\frac{E_g}{2} \nonumber \\
	& t_{dd}^{(1)} = \frac{1}{256} \left(-16 \widetilde{m}_{cx}^{-1}+16 \widetilde{m}_{cy}^{-1}+\frac{96 t_{dp}^2}{3 E_g-8 \widetilde{m}_{v1}^{-1}}-\frac{3 (8 \widetilde{m}_{v2}^{-1}-3 E_g) (2 \varepsilon_{c\Gamma}+E_g)^2}{(8 \widetilde{m}_{v2}^{-1}-3
   E_g) (2 \varepsilon_{c\Gamma}+E_g)+288 t_{dp}^2}+30 \varepsilon_{c\Gamma}-9 E_g\right) \nonumber \\
	& t_{pp\sigma} = -\frac{\widetilde{m}_{v1}^{-1}}{6} +\frac{\widetilde{m}_{v2}^{-1}}{2} + \frac{18 t_{dp}^2}{2 \varepsilon_{c\Gamma}+E_g} \label{Seq:fit_simple} \\
	& t_{pp\pi} = -\frac{\widetilde{m}_{v1}^{-1}}{2} + \frac{\widetilde{m}_{v2}^{-1}}{6} + \frac{6 t_{dp}^2}{2 \varepsilon_{c\Gamma}+E_g} \nonumber \\
	& t_{dd}^{(2)} = \frac{1}{256} \left(16 \widetilde{m}_{cx}^{-1}-16 \widetilde{m}_{cy}^{-1}+\frac{96 t_{dp}^2}{8 \widetilde{m}_{v1}^{-1}-3 E_g}-\frac{3 (3 E_g-8 \widetilde{m}_{v2}^{-1}) (2 \varepsilon_{c\Gamma}+E_g)^2}{(8 \widetilde{m}_{v2}^{-1}-3
   E_g) (2 \varepsilon_{c\Gamma}+E_g)+288 t_{dp}^2}+2 \varepsilon_{c\Gamma}-7 E_g\right) \nonumber \\	
	& t_{dd}^{(3)} = \frac{1}{1536} \left[ -48 \widetilde{m}_{cx}^{-1}-80 \widetilde{m}_{cy}^{-1}+3 \left(\frac{96 t_{dp}^2}{3 E_g-8 \widetilde{m}_{v1}^{-1}}-\frac{5 (3 E_g-8 \widetilde{m}_{v2}^{-1}) (2 \varepsilon_{c\Gamma}+E_g)^2}{(8 \widetilde{m}_{v2}^{-1}-3 E_g)
   (2 \varepsilon_{c\Gamma}+E_g)+288 t_{dp}^2}-2 \varepsilon_{c\Gamma}-9 E_g\right) \right] \nonumber
\end{align}
\end{widetext}

\begin{table*}[t]
\centering
\begin{tabularx}{1\linewidth}{LLLLLLLLL} 
 \hline \hline
 Reference & Technique & $m_{v1}/m_e$ & $m_{v2}/m_e$ & $m_{cy}/m_e$ & $m_{cx}/m_e$ & $m_{v2}/m_{v1}$ & $-m_{cy}/m_{v2}$ & $m_{cy}/m_{cx}$ \\ \hline
 \cite{kolekar_controlling_2018} & ARPES 2D & $-0.7$ & $-0.45$ & $7.1$ & $?$ & 0.64 & 16 & ? \\ 
 \cite{cercellier_evidence_2007,monney_spontaneous_2009} & ARPES 3D & $?$ & $-0.23$ & $5.5$ & $2.2$ & ? & 24 & 2.5 \\ 
 \cite{monney_dramatic_2009} & ARPES 3D & $?$  & $?$ & $6$ & $0.5$ & ? & ? & 12 \\ 
 \cite{chen_reproduction_2018} & DFT 2D & $?$ & $-0.19$ & $3.46$ & $0.22$ & ? & 18 & 16 \\ 
 \cite{guster_first_2018} & DFT 2D & $-0.25$ & $-0.15$ & $5.6$ & $0.4$ & 0.60 & 37 & 14 \\ 
 \cite{monney_impact_2015} & DFT 3D & $?$ & $-0.22$ & $4.3$ & $0.29$ & ? & 20 & 15 \\ \hline \hline
\end{tabularx}
\caption{Values of the masses of the valence and conduction bands obtained in previous works.}
\label{table:masses_fit}
\end{table*}

Table \ref{table:masses_fit} shows different values of the masses of the bands extracted from previous works. Here, we choose the values of Ref.~\cite{kolekar_controlling_2018} based on ARPES measurements on monolayer TiSe$_2$: $m_{v1} = -0.7 m_e$, $m_{v2} = (50/3) m_{v2} = -0.42 m_e$, $m_{cy} = 10 m_{v1} = 7 m_e$, $m_{cx} = m_{cy}/14 = 0.5 m_e$. Then, for gap $E_g = 0$ as in the inset of Fig.~\ref{Fig4} of the main text, the hopping parameters are $\varepsilon_d \simeq 0.329 \mathrm{eV}$, $\varepsilon_p \simeq -2.016 \mathrm{eV}$, $t_{dd}^{(1)} \simeq -0.017 \mathrm{eV}$, $t_{dd}^{(2)} \simeq 0.092 \mathrm{eV}$, $t_{dd}^{(3)} \simeq -0.030 \mathrm{eV}$, $t_{pp\sigma} \simeq -0.429 \mathrm{eV}$, $t_{pp\pi} \simeq 0.243 \mathrm{eV}$, and $t_{dp} \simeq 0.1 \mathrm{eV}$.

\subsection{Mean-field theory: order parameters \label{app:MF_order_parameters}}

\begin{figure}
    \centering
    \includegraphics[width=0.3\textwidth]{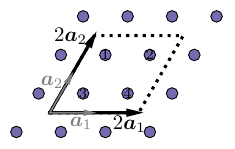}
    \caption{Triangular lattice of the effective tight-binding model with our choice of lattice vectors and supercell.}
    \label{Sfig:supercell_TB}
\end{figure}

By decoupling the Hamiltonian $H_0 + H_{\rm int}$ of Eqs.~(6)-(7) of the main text in the onsite orbital order $\langle d\dag_i \boldsymbol{p}_i \rangle$ channel, which corresponds to the Fock channel of the $V_{dp}$ interaction, the mean-field Hamiltonian in real-space becomes:
{\begin{align}
    H_{\rm MF} = H_0 + \sum_{\langle i j \rangle} & \left\{\left[\langle d_i^\dagger \boldsymbol{p}_i  \rangle^* \cdot \left(\hat{V}_{dp}\right)_{ij} \cdot  \left( d_j^\dagger \boldsymbol{p}_j \right) + \mathrm{h.c.}\right] - \right.\nonumber\\
    &\left.-\langle d_i^\dagger \boldsymbol{p}_i  \rangle^* \cdot \left(\hat{V}_{dp}\right)_{ij} \cdot \langle d_j^\dagger \boldsymbol{p}_j \rangle \right\},
\label{Seq:H_MF}
\end{align}
where the expectation values are taken in the mean-field ground state and $\left(\hat{V}_{dp}\right)_{ij}= 2 \hat{\boldsymbol{r}}_{ij} \otimes \hat{\boldsymbol{r}}_{ij} - 1 $.} We solve $H_{\rm MF}$ in a $2\times2$ supercell with superlattice vectors $2\boldsymbol{a}_i$, with $\boldsymbol{a}_1 = a(1,0)$ and $\boldsymbol{a}_2 = a(\tfrac{1}{2},\tfrac{\sqrt{3}}{2})$ and label the supercell sites by $j=1,2,3,4$ (see Fig.~\ref{Sfig:supercell_TB}). Without loss of generality, we choose the origin of coordinates of a cell at $\boldsymbol{r}_1=(0,0)$, so $\boldsymbol{r}_2=a(1,0)$, $\boldsymbol{r}_3=a(-1/2,-\sqrt{3}/2)$, and $\boldsymbol{r}_4=a(1/2,-\sqrt{3}/2)$.

Let us classify according to symmetry the different terms entering the Hamiltonian. In the original unit cell, the onsite orbital orders $(d\dag p_x \pm \mathrm{h.c.}, d\dag p_y \pm \mathrm{h.c.})$ transform as TRS even and odd $\Gamma_3^-$, which correspond to the irrep $E_{u}$.

Now, let us consider the symmetry classification in the $2\times2$ supercell. The orbital orders $d\dag p_{\alpha}$ transform as $\Gamma_3^- \oplus M_{1}^- \oplus M_{2}^-$, 
where the $M_{1}^-$ and $M_{2}^-$ are irreps of the little group of the $M$ points that are even and odd under the $C_2$ symmetry, respectively. The symmetry adapted operators read
\begin{align}
    & \left[d\dag p\right]_{\Gamma_3^-}^{\alpha} = \frac{1}{2} \sum_{j} d_j\dag p_{\alpha j}, \label{Seq:OP_short_ini} \\
    & \left[d\dag p\right]_{M_{1}^-}^a = \frac{1}{2} \sum_{j} e^{i \boldsymbol{M}_a \cdot \boldsymbol{r}_j} d_j\dag \boldsymbol{p}_j \wedge \hat{\boldsymbol{M}}_a, \label{Seq:OP_short_middle} \\ 
    & \left[d\dag p\right]_{M_{2}^-}^a = \frac{1}{2} \sum_{j} e^{i \boldsymbol{M}_a \cdot \boldsymbol{r}_j} d_j\dag \boldsymbol{p}_j \cdot \hat{\boldsymbol{M}}_a, 
\label{Seq:OP_short_fin}
\end{align}
where the subindices label the irrep according to which the operators transform, and the superindices $\alpha=1,2$ and $a=1,2,3$ label the different components of the multidimensional irreps. We have defined $\hat{\boldsymbol{M}}_a = \frac{\boldsymbol{M}_a}{|\boldsymbol{M}_a|}$, and $\boldsymbol{v} \wedge \boldsymbol{w} = v_x w_y - v_y w_x$. We use the notation indicated in Fig.~\ref{Fig1} of the main text: $\boldsymbol{M}_1 = \tfrac{2\pi}{\sqrt{3}a}(0,1)$, $\boldsymbol{M}_2 = \tfrac{2\pi}{\sqrt{3}a}(-\tfrac{\sqrt{3}}{2},-\tfrac{1}{2})$, $\boldsymbol{M}_3 = \tfrac{2\pi}{\sqrt{3}a}(\tfrac{\sqrt{3}}{2},-\tfrac{1}{2})$. TRS even and odd operators can be obtained by adding or substracting the Hermitian conjugate, respectively. This implies that the real and imaginary parts of the $\langle d\dag p\rangle$ expectation values are even and odd under TRS, respectively.

Using the previous symmetry-adapted operators and the mean-field interaction of Eq.~\eqref{Seq:H_MF}, let us construct the symmetry-adapted order parameter $\vec{\Delta}$ that couples the valence and conduction bands as in Eqs. (\ref{eq:H_kp}-\ref{eq:HDelta}) of the main text at low energies. {For the mean-field $\langle d\dag p \rangle$ decoupling in the $2\times2$ supercell, we can regard $\hat{V}_{dp}$ as a matrix in the basis $\{ d_1\dag p_{x1} , d_1\dag p_{y1} , d_2\dag p_{x2} , d_2\dag p_{y2} , d_3\dag p_{x3} , d_3\dag p_{y3} , d_4\dag p_{x4} , d_4\dag p_{y4} \}$:
\begin{align}
    H_{\rm MF} = H_0 + \sum_{I} & \left\{\left[\langle d^\dagger \boldsymbol{p}  \rangle^* \cdot \hat{V}_{dp} \cdot  \left( d_{I}^\dagger \boldsymbol{p}_{I} \right) + \mathrm{h.c.}\right] -\right. \nonumber \\
    &-\left.\langle d^\dagger \boldsymbol{p}  \rangle^* \cdot \hat{V}_{dp} \cdot \langle d^\dagger \boldsymbol{p} \rangle \right\},
\label{Seq:H_MF2}
\end{align}
where $I$ labels the supercells and:
{\small \begin{equation}
    \frac{\hat{V}_{dp}}{V_{dp}} = 
    \begin{pmatrix}
    0    & 0    & 1    & 0    & -\tfrac{1}{2} & \tfrac{\sqrt{3}}{2}  & -\tfrac{1}{2} & -\tfrac{\sqrt{3}}{2} \\
    0    & 0    & 0    & -1   & \tfrac{\sqrt{3}}{2}  & \tfrac{1}{2}  & -\tfrac{\sqrt{3}}{2} & \tfrac{1}{2}  \\
    1    & 0    & 0    & 0    & -\tfrac{1}{2} & -\tfrac{\sqrt{3}}{2} & -\tfrac{1}{2} & \tfrac{\sqrt{3}}{2}  \\
    0    & -1   & 0    & 0    & -\tfrac{\sqrt{3}}{2} & \tfrac{1}{2}  & \tfrac{\sqrt{3}}{2}  & \tfrac{1}{2}  \\
    -\tfrac{1}{2} & \tfrac{\sqrt{3}}{2}  & -\tfrac{1}{2} & -\tfrac{\sqrt{3}}{2} & 0    & 0    & 1    & 0    \\
    \tfrac{\sqrt{3}}{2}  & \tfrac{1}{2}  & -\tfrac{\sqrt{3}}{2} & \tfrac{1}{2}  & 0    & 0    & 0    & -1   \\
    -\tfrac{1}{2} & -\tfrac{\sqrt{3}}{2} & -\tfrac{1}{2} & \tfrac{\sqrt{3}}{2}  & 1    & 0    & 0    & 0    \\
    -\tfrac{\sqrt{3}}{2} & \tfrac{1}{2}  & \tfrac{\sqrt{3}}{2}  & \tfrac{1}{2}  & 0    & -1   & 0    & 0   
    \end{pmatrix}.
\end{equation}}}
Diagonalizing this matrix, we get eigenvalue $-2 V_{dp}$ for the eigenvector $\left[d\dag p\right]_{M_{1}^-}$ \eqref{Seq:OP_short_middle}, $+2 V_{dp}$ for $\left[d\dag p\right]_{M_{2}^-}$ \eqref{Seq:OP_short_fin}, and $0$ for $\left[d\dag p\right]_{\Gamma_3^-}$ \eqref{Seq:OP_short_fin}. {Consequently, we can recast $H_{\rm MF}$ in Eq. \eqref{Seq:H_MF} as 
{\small \begin{equation}
\begin{split}
    \frac{H_{\rm MF}}{2 V_{dp}} = \sum_{I} & \left\{- \left( \langle d\dag p \rangle^*_{M_{1}^-} \left[d_I\dag p_I\right]_{M_{1}^-} + \mathrm{h.c.} - \big| \langle d\dag p \rangle_{M_{1}^-} \big|^2 \right) + \right. \\
    & \left. + \left( \langle d\dag p \rangle^*_{M_{2}^-} \left[d_I\dag p_I\right]_{M_{2}^-} + \mathrm{h.c.} - \big| \langle d\dag p \rangle_{M_{2}^-} \big|^2 \right) \right\}.
\end{split}
\label{Seq:Hdp1_MF_irreps}
\end{equation}}
As we anticipated, the mean-field Hamiltonian~\eqref{Seq:H_MF} is only attractive in the $M_1^-$ channel.

The last step to write the explicit expression of the order parameter $\vec{\Delta}$ is to determine its normalization so that it couples to the valence and conduction bands in the $\boldsymbol{k}\cdot\boldsymbol{p}$ model at $\boldsymbol{k}=0$ as in $H_{dp}(\vec{\Delta})$ of Eqs.~(\ref{eq:H_kp},\ref{eq:HDelta}) of the main text. For that, we express the low-energy $\boldsymbol{k}\cdot\boldsymbol{p}$ eigenfunctions $\{d_1^{kp}, d_2^{kp}, d_3^{kp}, p_x^{kp}, p_y^{kp}\}$ in the basis of tight-binding orbitals $\{d_j, p_{xj}, p_{yj}\}_{j=1}^4$ in the $2\times 2$ supercell. Taking into account that $d_a^{kp}$ have momentum $\Gamma M$ and $p_\alpha^{kp}$ have zero momentum, and that the eigenfunctions at $\Gamma$ and $M$ are purely $p$ and purely $d$ in the tight binding, respectively, we obtain:
\begin{align}
    & d_a^{kp} = \frac{1}{2} \sum_j e^{i \boldsymbol{M}_a \cdot \boldsymbol{r}_j} d_j, \\
    & p_\alpha^{kp} = \frac{1}{2} \sum_j p_\alpha,
\end{align}
The projection of the time-reversal symmetric $M_1^-$ component of $H_{dp1}^{\rm MF}$ onto the $\boldsymbol{k}\cdot\boldsymbol{p}$ bands at momentum $\boldsymbol{k}=0$ of the CDW Brillouin zone is therefore as in Eq.~\eqref{eq:H_kp} of the main text with
{\small \begin{align}
    H_{dp}^\dagger = \frac{\left( -2 V_{dp} \right)}{2} \mathrm{Re} \left(\begin{array}{ccc}
    \langle d\dag p \rangle_{M_{1}^-}^1 &  -\tfrac{1}{2} \langle d\dag p \rangle_{M_{1}^-}^2 & -\tfrac{1}{2} \langle d\dag p \rangle_{M_{1}^-}^3 \\
    0 & \tfrac{\sqrt{3}}{2} \langle d\dag p \rangle_{M_{1}^-}^2 &  -\tfrac{\sqrt{3}}{2} \langle d\dag p \rangle_{M_{1}^-}^3 \\
    \end{array}
    \right).
    \label{Seq:Hdp_kp_TB}
\end{align}}
Comparing this expression to Eq.~\eqref{eq:HDelta}, we conclude that the explicit expression of the time-reversal symmetric $M_1^-$ CDW order parameter $\vec{\Delta}$ in the tight binding is 
\begin{align}
	\Delta_a &= \frac{1}{2} \left( -2 V_{dp} \right) \mathrm{Re} \langle d_j\dag \boldsymbol{p}_j \rangle_{M_1^-}^a \nonumber\\
    &= - {\frac{1}{2}} V_{dp} \sum_{j} e^{i \boldsymbol{M}_a \cdot \boldsymbol{r}_j} \mathrm{Re} \langle d_j\dag \boldsymbol{p}_j \rangle \wedge \hat{\boldsymbol{M}}_a,
    \label{Seq:Delta}
\end{align}}

Our mean-field decoupling \eqref{Seq:H_MF} allows for another three order parameters. One is the TRS odd $M_{1}^-$ counterpart of $\vec{\Delta}$, which corresponds to taking the imaginary part instead of the real part in Eq.~\eqref{Seq:Delta}. In the presence of separate charge conservation for $p$ and $d$ orbitals, the time-reversal even and odd $M_{1}^-$ order parameters would be degenerate~\cite{ganesh_theoretical_2014}. However, the presence of a small $t_{dp}$ hopping breaks their degeneracy in favor of the real part $\vec{\Delta}$ in all cases. Indeed, the TRS odd $M_{1}^-$ order parameter is identically zero in our calculations. The other two order parameters transform as TRS even and odd $M_{2}^-$, ${\Psi_{a\pm}} = + {\frac{1}{2}} V_{dp} \sum_{j} \tfrac{1}{2}( e^{i \boldsymbol{M}_a \cdot \boldsymbol{r}_j} \langle d_j\dag \boldsymbol{p}_j \rangle \cdot \hat{\boldsymbol{M}}_a \pm \mathrm{h.c.})$, and the interaction is repulsive for both. {$\vec{\Psi}_+$ is nonzero but small only in the nematic phase, where it is symmetry allowed, with $\vec{\Delta} = (\Delta_1,\Delta_2,\Delta_2)$ and $\vec{\Psi}_+ = (0,\Psi_{+},-\Psi_{+})$. Therefore, our interacting tight-binding model only favors time-reversal even $M_{1}^-$ instabilities, as required. Our interacting tight-binding Hamiltonian therefore serves as a minimal model to analyze the energetics of the $M_1^-$ CDW order.}

\subsection{Mean-field theory: calculation details}

We perform self-consistent mean-field calculations on the Hamiltonian of Eq.~\eqref{Seq:H_MF}. For that, we introduce a initial seed for the order parameters, and recompute them iteratively until convergence is reached, defined as $\sqrt{\sum |\langle c\dag c \rangle_{n+1} - \langle c\dag c \rangle_{n}|} < \epsilon_0$. Since we work in the canonical ensemble, in each iteration we set the chemical potential to keep the number of particles fixed. 

In order to find the ground state which minimizes the free energy, we initialize the self-consistent loop with different seeds: $\vec{\Delta} = (\Delta,\Delta,\Delta)$, $\vec{\Delta} = (\Delta,0,0)$, $\vec{\Delta} = (0,\Delta,\Delta)$, $\vec{\Delta} = (\Delta_1,\Delta_2,\Delta_2)$, and $\vec{\Delta} = (\Delta_1,\Delta_2,\Delta_3)$. To guarantee that each seed converges to the phase that we want, we first run the self-consistent loop by symmetry-restricting the mean-field parameters to have the symmetry they initially have. Once convergence has been reached, we run unrestricted self-consistent loops whose seeds are the solutions of the restricted loops.

\subsection{Additional \textit{T} -- \textit{x} phase diagrams}

\begin{figure*}[t]
    \centering
    \includegraphics[width=\textwidth]{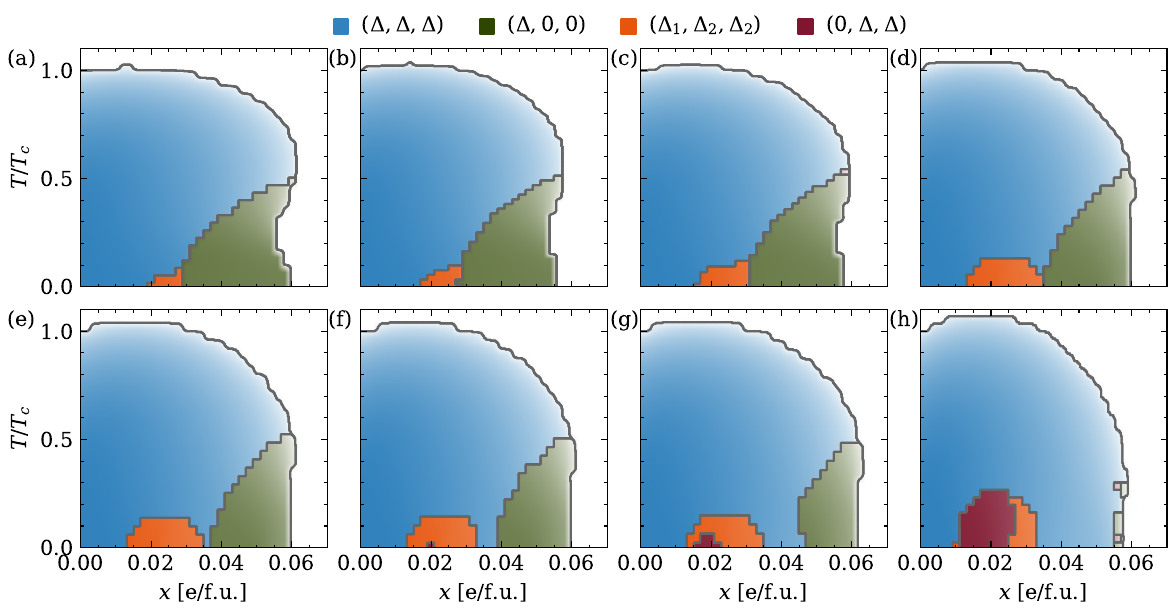}
    \caption{Temperature-doping phase diagrams obtained by self-consistently solving the mean-field Hamiltonian of Eq.~\eqref{Seq:H_MF}:\\ 
    (a) $E_g = +50 \mathrm{meV}$, $V_{dp} = 890 \mathrm{meV}$, $T_c = 576\mathrm{K}$;\\
    (b) $E_g = +25 \mathrm{meV}$, $V_{dp} = 845 \mathrm{meV}$, $T_c = 459\mathrm{K}$;\\
    (c) $E_g = 0$, $V_{dp} = 800 \mathrm{meV}$, $T_c = 378\mathrm{K}$;\\
    (d) $E_g = -25 \mathrm{meV}$, $V_{dp} = 740 \mathrm{meV}$, $T_c = 269\mathrm{K}$;\\
    (e) $E_g = -27.5 \mathrm{meV}$, $V_{dp} = 733 \mathrm{meV}$, $T_c = 259\mathrm{K}$;\\
    (f) $E_g = -30 \mathrm{meV}$, $V_{dp} = 725 \mathrm{meV}$, $T_c = 249\mathrm{K}$;\\
    (g) $E_g = -35 \mathrm{meV}$, $V_{dp} = 716.5 \mathrm{meV}$, $T_c = 239\mathrm{K}$;\\
    (h) $E_g = -50 \mathrm{meV}$, $V_{dp} = 650 \mathrm{meV}$, $T_c = 145\mathrm{K}$.}
    \label{SFig:more_phase_diagrams}
\end{figure*}

\begin{figure}[t]
    \centering
    \includegraphics[width=\linewidth]{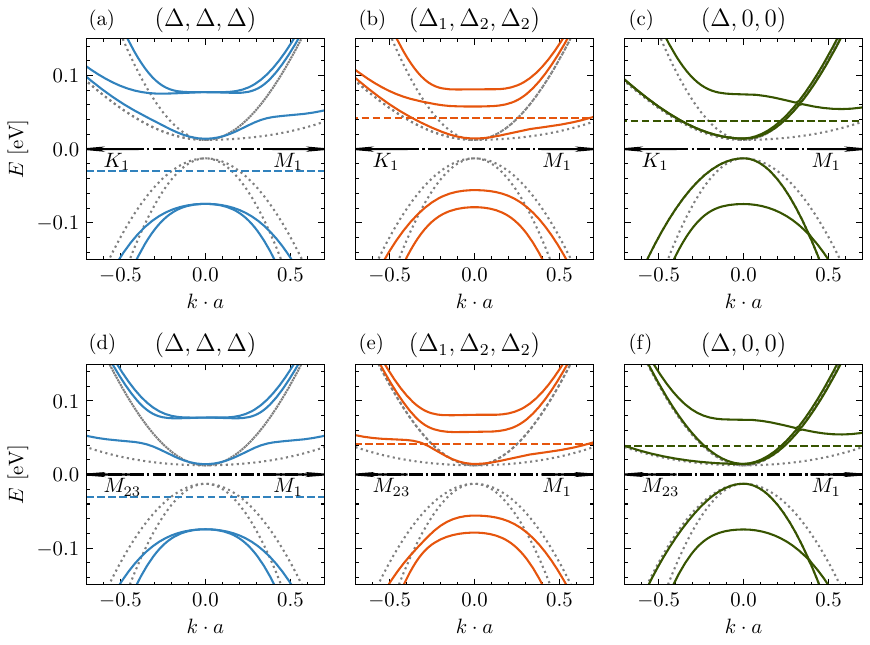}
    \caption{{Band structure along the $K_1 \Gamma M_1$ (top row) and $M_{23} \Gamma M_1$ (bottom row) directions for the zero-temperature ground state of the effective lattice model for $E_g=+25\mathrm{meV}$ and $V_{dp} = 845\mathrm{meV}$ at different dopings (the corresponding chemical potential is shown as a dashed line). Dotted grey lines represent the band structure of the normal state without CDW.
    (a,d) $x=0 \mathrm{e/f.u.}$, where the ground state is the $C_3$-symmetric $3Q$ state with $|\vec{\Delta}| \simeq {172} \mathrm{meV}$. 
    (b,e) $x=0.022 \mathrm{e/f.u.}$, where the ground state is the $3Q$ nematic state $\vec{\Delta}=(\Delta_1,\Delta_2,\Delta_2)$ with $|\vec{\Delta}| \simeq {157} \mathrm{meV}$ and $|\Delta_1/\Delta_2| \simeq 0.51$.
    (c,f) $x=0.04 \mathrm{e/f.u.}$, where the ground state is the $1Q$ stripe state with $|\vec{\Delta}| \simeq {120} \mathrm{meV}$.}} 
    \label{SFig:bands_TB}
\end{figure}

In this section we discuss phase diagrams for different gaps and interaction strengths. The main goal of this section is to provide further support to the claim that the existence of $C_3$-breaking phases is qualitatively robust.

In our simplified effective model with a given initial gap $E_g$, it is not possible to choose a value of the interaction that reproduces both the critical temperature $T_c$ and the critical doping $x_c$ for the disappearance of the CDW. Our aim is not to reproduce these values quantitatively (which would require including physics well beyond our model like electron-phonon coupling~\cite{zhou_anharmonicity_2020}, doping-dependent screening, the effect of the substrate~\cite{li_enhancing_2016,kolekar_controlling_2018}, fluctuation corrections to mean field~\cite{kos_gaussian_2004} \ldots), but rather to show the generic existence of $C_3$-breaking phases. Because of this, we choose interaction strengths to match the critical doping above which the commensurate CDW dies and present the phase diagrams as a function of $T/T_c$, noting that $T_c$ generally changes for different values of $E_g$. The critical doping we take is $x_c \sim 0.06 \mathrm{e/f.u.}$, as seen experimentally~\cite{wu_transport_2007,LOL15,KogarICDW,Watson20}. This is an approximate estimate, since at this doping there is a crossover to an incommensurate CDW, where the ground state consists of commensurate domains separated by domain walls~\cite{mcmillan_landau_1975,mcmillan_theory_1976,KogarICDW,Yan17}.

Taking this into account, Fig.~\ref{SFig:more_phase_diagrams} shows how the $T$-$x$ phase diagrams evolve from gap $E_g = +50 \mathrm{meV}$ to $E_g = -50 \mathrm{meV}$. The most prominent feature is the shift of the $1Q$ stripe phase to higher doping for decreasing gap, until this phase disappears. This is consistent with the $\boldsymbol{k} \cdot \boldsymbol{p}$ predictions of Figs.~\ref{Fig2}(e,f), which show that the doping above which the $1Q$ solution is more stable than the $3Q$ increases with decreasing gap. Furthermore, while the $3Q$ nematic and the $1Q$ stripe phases share a boundary for positive gaps, they separate for negative gaps, clearly indicating that the origin of these two phases is different, which agrees with our theory. Finally, the relative extension of the $2Q$ phase with respect to the $3Q$ nematic phase increases with decreasing gap.

Finally, in Fig.~\ref{SFig:bands_TB} we plot the low-energy band structure of the effective lattice model in the CDW state for $E_g = +25\mathrm{meV}$ three different dopings where the zero temperature ground state corresponds to the $3Q$ $C_3$-symmetric, $3Q$ nematic and $1Q$ stripe states. The first row shows the dispersion along the $K_1 \Gamma M_1$ direction, and can be directly compared to Figs.~\ref{Fig2}(a-c) of the main text, which show the corresponding $\boldsymbol{k}\cdot\boldsymbol{p}$ band structures. The second row shows the dispersion along the $M_{23} \Gamma M_1$ direction, highlighting the breaking of $C_3$ symmetry in the nematic and stripe band structure.

\bibliography{TiSe2}

\end{document}